\documentclass[a4paper,11pt]{article} 
\pdfoutput=1 
\usepackage{jheppub} 
\usepackage{graphicx,hyperref,xcolor,amsmath,amssymb,bm}
\allowdisplaybreaks
\newcommand{\as}{\alpha_s} 
\newcommand{\asc}{\frac{\alpha_s C_F}{2\pi}}
\newcommand{\acz}{\frac{\alpha_s C_F}{2\pi}(4\zeta)}
\newcommand{\eps}{\epsilon} 
\newcommand{\euv}{\eps_{\mathrm{UV}}}
\newcommand{\eir}{\eps_{\mathrm{IR}}}
 \newcommand{\fms}[1]{{#1}\!\!\!/}

\newcommand{\msbar}{\overline{\text{MS}}}

\newcommand{\ribar}{\overline{\text{RI/MOM}}}
\newcommand{\rbar}{\overline{\text{RI}}}

\newcommand{\mRI}{\mathrm{mRI}}

\begin{document}

\title{Isolating Scheme Dependence of Quasi-PDFs in the RI/MOM  Scheme}
\def\KU{Department of Physics, Korea University, Seoul 02841, Korea} 
\author[]{Junegone Chay}
\emailAdd{chay@korea.ac.kr}
\affiliation[ ]{\KU}
 
\abstract{
Quasi-parton distribution functions provide a framework for relating lightcone parton distributions to correlation 
functions in Euclidean lattice QCD. Their connection to lightcone distributions is established through perturbative matching, 
whose form depends on the renormalization prescription adopted for the quasi-PDF. In the ordinary RI/MOM scheme, the 
off-shell reference momentum enters both the renormalized quasi-PDF and the matching coefficient. We propose modified 
finite renormalization schemes in which neither quantity depends on the RI/MOM reference momentum. Starting from the 
ordinary RI/MOM scheme, we identify the part to be retained in the renormalized quasi-PDF and include the remaining finite 
part in the counterterm. We consider a minimal RI/MOM scheme and a modified minimal scheme as specific examples. We 
derive the corresponding one-loop renormalized quasi-PDFs, counterterms, matching coefficients, and renormalization-group
equations, and show explicitly that the finite transformation removes the dependence on the RI/MOM reference 
momentum from both the renormalized quasi-PDF and the counterterm in the modified schemes.}

\maketitle

\section{Introduction}

High-energy hadronic processes involve physics at widely separated momentum scales. Factorization separates the 
short-distance part, which can be computed perturbatively in QCD, from long-distance hadronic quantities that
contain nonperturbative information. Parton distribution functions (PDFs) are among the most important of these 
quantities. They describe the longitudinal momentum distributions of partons inside a hadron and enter the factorized 
description in a broad class of hard-scattering processes. Their scale dependence is governed perturbatively, but their values 
at a given scale must be determined from nonperturbative information. They are therefore extracted phenomenologically 
from experimental data or studied through calculations from first principles in lattice QCD.

The direct calculation of PDFs in Euclidean lattice QCD is obstructed by their definition in terms of lightcone correlations. 
Large-Momentum Effective Theory (LaMET)~\cite{Ji:2013dva,Ji:2014gla} mitigates this difficulty by introducing quasi-parton 
distribution functions (quasi-PDFs), which are defined from equal-time spatial correlations in a hadron carrying a large 
momentum and are therefore accessible in lattice QCD. Since the advent of LaMET, there has been an intensive effort to 
formulate and refine this relation~\cite{Ji:2017rah}.  For reviews, see Refs.~\cite{Ji:2020ect,Cichy:2018mum}. 
The relation between the PDF $q$ and the quasi-PDF $\tilde q$ takes the factorized form~\cite{Ma:2014jga,Izubuchi:2018srq}, 
and is schematically written as
\begin{equation}
\tilde{q} = C \otimes q,
\end{equation}
where $C$ is the matching coefficient and $\otimes$ stands for an appropriate convolution. The one-loop matching for 
nonsinglet quasi-PDFs was first derived in a transverse-momentum cutoff scheme in Ref.~\cite{Xiong:2013bka}.
However, this constitutes only one step in relating the lightcone PDF to the
spatial correlation accessible in lattice QCD. A second matching relates the
quasi-PDF to the lattice correlation, so that the complete construction
proceeds through two successive matching relations.

 In computing the matching coefficient in any stage, we should employ a renormalization scheme. In QCD, the $\msbar$ scheme 
is accepted as a standard one. The ultraviolet (UV) and infrared (IR) divergences are handled 
by dimensional regularization with spacetime dimension $D = 4-2\eps$. Sometimes the IR divergence is 
treated by the off-shellness of external states. A renormalization scheme is needed for quasi-PDFs as well, and
any renormalization prescription unavoidably introduces scheme dependence. The scheme dependence affects the 
renormalized quasi-PDF and the matching coefficient accordingly. The apparent choice of the renormalization scheme for 
the quasi-PDF to match onto the PDF would be to use the same $\msbar$ scheme as in the PDF. It is conceptually 
straightforward, though the calculations are still involved~\cite{Izubuchi:2018srq}. It is a legitimate choice, but in dimensional
regularization only logarithmic divergences appear as poles in $1/\eps$, while linear and other power divergences are set to 
zero.  On the other hand, in lattice QCD, we know that the spatial 
correlator contains a linear divergence, which comes from the self-energy of the Wilson lines in some schemes. The 
renormalization of straight Wilson-line operators and the associated linear divergence have been studied perturbatively,
through Wilson-line mass-renormalization methods, and through all-order multiplicative renormalizability arguments
~\cite{Ji:2015jwa,Chen:2016fxx,Constantinou:2017sej,Ji:2017oey, Li:2018tpe,Ishikawa:2017faj}. Therefore the 
treatment of linear divergence is deferred to the matching between the quasi-PDF in the $\msbar$ scheme and the 
correlation in lattice QCD in this case, but we will not treat the matching involving lattice QCD here.

Several renormalization prescriptions for quasi-PDFs have instead been developed from the perspective of lattice QCD. 
Examples include the regularization-independent momentum-subtraction (RI/MOM) scheme~\cite{Martinelli:1994ty,Alexandrou:2017huk}, the ratio scheme~\cite{Orginos:2017kos,Radyushkin:2017cyf}, and the 
hybrid scheme~\cite{Ji:2020brr,Chou:2022drv}, each motivated by different theoretical or practical considerations. These 
prescriptions are designed to renormalize the spatial correlations calculated in lattice QCD. It is then convenient to use a 
compatible prescription for the quasi-PDF, just as the $\msbar$ scheme is used consistently in matching the quasi-PDF 
onto the lightcone PDF. The different approaches vary in how the renormalization prescription is shared between the lattice
 correlation and the quasi-PDF.

In matching quasi-PDFs to correlations in lattice QCD using these schemes, the resulting expressions, however, often 
intertwine the universal collinear structure with contributions tied to the particular renormalization prescription.  This can 
make it difficult to compare the results in different schemes and to understand the physics contained in the renormalized 
quasi-PDF and the matching coefficient. Our motivation is therefore to seek a method in which the contribution 
associated with the renormalization prescription is clarified and separated as transparently as possible from the part required 
by collinear factorization.

Following this line of thought, we studied the isolation of scheme dependence in the transverse-momentum cutoff (TMC) 
scheme for quasi-PDFs in Ref.~\cite{Chay:2026cyw}. The TMC scheme is not commonly used as a practical renormalization 
prescription in lattice QCD, but it has the advantage that the cutoff dependence is revealed explicitly.  The linear divergence 
from the self-energy of the Wilson line and the logarithmic cutoff dependence can be identified directly~\footnote{The 
one-loop gluon quasi-PDF was also calculated with a transverse-momentum cutoff, and dimensional regularization 
in Ref.~\cite{Wang:2017qyg}, in which the authors find that, unlike the quark quasi-PDF, the gluon quasi-PDF contains a 
linear divergence arising from a diagram that is not connected to the Wilson line.}. The bare quasi-PDF can be decomposed 
into a counterterm that depends on the cutoff and a renormalized quasi-PDF that is independent of the cutoff.  The renormalized 
quasi-PDF is then matched onto the lightcone PDF after the linear divergence in the TMC scheme is properly taken care of.  In 
that setting the separation of scheme dependence is relatively simple.  However, the TMC analysis poses a more general question. 
Which part of the quasi-PDF belongs to the renormalization prescription, and which part is essential for collinear matching 
onto the PDF? 
  
Compared to the TMC scheme, the RI/MOM  scheme  provides a more subtle testing ground for these questions. In the 
RI/MOM scheme, the scheme dependence is not tied to an explicit cutoff. Instead, it is associated with the reference 
momentum at which the RI/MOM renormalization condition is imposed. From now on, we call the RI/MOM scheme in 
Ref.~\cite{Stewart:2017tvs}  the ordinary RI/MOM scheme in order to distinguish it from the modified RI/MOM schemes 
to be introduced in this paper.  We start from the one-loop quasi-PDF in the ordinary RI/MOM scheme as a concrete setting 
in which the dependence on the RI/MOM reference momentum appears explicitly. We then ask whether we can find 
appropriate prescriptions in which this dependence on the reference momentum can be removed from the renormalized 
quasi-PDF and the counterterm in a controllable way.

In the ordinary RI/MOM scheme, a renormalization prescription at a specific reference momentum determines the 
renormalized quasi-PDF, the counterterm, and eventually the matching coefficient. The dependence on this reference momentum, 
however, pervades all of them. Although this causes no problem in the definition of the scheme, it can obscure which terms 
arise from the choice of renormalization scheme and which terms encode the collinear structure required for matching to the 
lightcone PDF. This motivates asking whether all the dependence on the reference momentum can be removed from
the renormalized quasi-PDF so that matching becomes  more transparent. To achieve this separation, we construct finite 
scheme transformations and trace their consequences for the renormalized quasi-PDF and the matching coefficient. These 
are the questions addressed here.

The identification and separation of the scheme dependence, however, are not arbitrary. For the matching to be consistent, 
the renormalized 
quasi-PDF must retain the collinear structure shared with the lightcone PDF, so that the matching coefficient is infrared safe. 
We also require the separation to satisfy quark-number conservation. This is an additional requirement for defining the 
modified schemes rather than a condition essential for perturbative matching itself. These requirements do not uniquely 
determine the remaining finite terms. Some of them may be retained in the renormalized quasi-PDF or assigned to the 
counterterm, and this freedom gives rise to different renormalization schemes.

The RI/MOM reference momentum remains essential for the nonperturbative renormalization of the spatial Wilson-line 
operator in lattice QCD~\cite{Constantinou:2017sej,Chen:2017mzz,Alexandrou:2017huk}, and it can be employed in the 
quasi-PDF to facilitate matching between the correlations calculated in lattice QCD and the quasi-PDF.  Our question 
concerns the subsequent perturbative matching between the quasi-PDF and the PDF.  Starting from the quasi-PDF in the 
ordinary RI/MOM scheme, can the contribution tied to the reference momentum be confined to the renormalization factor 
while the quasi-PDF used for matching retains the required collinear structure and preserves quark-number conservation? 
To answer this question, we formulate a class of finite RI/MOM-type schemes and construct two explicit realizations, which 
we call the minimal RI/MOM and modified minimal RI/MOM, or $\ribar$, schemes.

We do not merely rewrite the result in the ordinary RI/MOM scheme by an algebraic rearrangement. Once a contribution 
satisfying the preceding conditions is chosen to be retained in the renormalized quasi-PDF, the corresponding counterterm 
and matching coefficient are fixed. Different choices for the remaining finite terms therefore define finite RI/MOM-type
renormalization schemes. Because these transformations may involve the RI/MOM reference scale, they can also change the 
anomalous dimensions of the quasi-PDF and the matching coefficient. Their scale dependences must therefore be transformed 
consistently with the scheme, as we discuss below.
  
Although the quasi-PDF can be defined as the spatial correlator in coordinate space, the separation of scheme dependence
is more transparent after Fourier transformation to the momentum-fraction variable $x$.  In $x$ space, the coefficient
of the collinear logarithm, the remaining finite terms, and the part depending on the reference momentum can be identified
separately, including their endpoint prescriptions.  This is the main reason for formulating the scheme separation in
$x$ space. In this representation, the common contribution required by collinear matching can be distinguished from the 
finite terms that remain to be fixed by the choice of scheme. The corresponding renormalized quasi-PDFs, counterterms, and 
matching coefficients can then be derived systematically.

The paper is organized as follows.  In Section~\ref{rev} we delineate the quasi-PDF in the ordinary RI/MOM scheme 
and introduce the notations to be used throughout the paper. In Section~\ref{sep}, we review the momentum-space 
quasi-PDF in the ordinary RI/MOM scheme and recast it into a form suitable for separating the contribution of the reference
momentum. In Section~\ref{minmom} we define and explain the minimal and modified minimal RI/MOM schemes.  
In Section~\ref{matco} we derive the corresponding matching coefficients, complete the results as distributions on the full 
line by including the boundary terms at $x=\pm\infty$, and discuss their renormalization-group behavior.
In Section~\ref{coint} we explain the implications of the scheme separation in coordinate space. We especially focus on 
where the spatial logarithms reside in each scheme and explain their implications. In Section~\ref{etagen} we extend the 
construction to the general case, where the external momentum $p^z$ and the momentum used in the
renormalization condition $p_R^z$ are not equal. In Section~\ref{sec:hybrid-comparison}, we compare our
construction with the hybrid-RI/MOM prescription and explain how the hybrid-ratio prescription is recovered when the 
longitudinal reference momentum is set to zero. Section~\ref{conc} contains our conclusions. In Appendix, we collect the 
prescriptions and expressions for distributions on the full line.
 
\section{Quasi-PDF in the RI/MOM scheme} \label{rev}

We reconstruct the result in the ordinary RI/MOM scheme in a form that makes explicit the separation between the bare 
quasi-PDF and the counterterm defined at the RI/MOM reference momentum. This form will provide the input for the finite
scheme construction in Sections~\ref{sep} and \ref{minmom}. We review the definitions of the lightcone PDF and the 
quasi-PDF and summarize the quasi-PDF in the ordinary RI/MOM scheme. This establishes the notation and the form of the 
result in momentum space needed for the subsequent analysis. We begin with the lightcone PDF, to which the quasi-PDF is 
perturbatively matched.

The quark PDF is defined as the lightcone correlation of the bilinear quark operator. Using   
dimensional regularization with $D= 4-2\eps$, the bare unpolarized nonsinglet quark PDF is 
\begin{equation}
q(x,\eps)  = \int \frac{d\xi^-}{4\pi} e^{ixP^+ \xi^-} \langle P |\overline{\psi} (\xi^- ) \gamma^+ W(\xi^-, 0)
\psi (0) |P\rangle,
\end{equation}
where $x$ is the longitudinal momentum fraction, $P^{\mu} = (P^0, 0, 0, P^z)$ is the nucleon momentum, and
$\xi^{\pm} = (t\pm z)/\sqrt{2}$ are the lightcone coordinates. The operator is gauge invariant due to the presence 
of the  Wilson line $W$, which is  
\begin{equation}
W(\xi^- ,0) = P \exp \left( -ig \int_0^{\xi^-} d\eta^- A^+ (\eta^-) \right),
\end{equation}
and $P$ denotes path ordering. The renormalized PDF $q(y,\mu)$ is defined in the $\msbar$ scheme as
\begin{equation}
q_B (x,\eps) = \int_x^1 \frac{dy}{y} Z_{\msbar} \Bigl( \frac{y}{x}, \eps, \mu\Bigr) q_R (y,\mu),
\end{equation} 
where $q_B$ ($q_R$) is the bare (renormalized) PDF, $Z_{\msbar}$ is the counterterm, and $\mu$ is the 
renormalization scale.  
 
The bare quasi-PDF, which is the spatial correlation function, is defined in coordinate space as
\begin{equation}
\tilde{Q} (z, P^z, \eps) \equiv  \langle P|\overline{\psi} (z) \frac{\gamma^z}{2} W_z (z,0) \psi (0)|P\rangle,
\end{equation}
and its Fourier transform is given by
\begin{equation} \label{hadfour}
\tilde{q} (x,P^z, \eps) =P^z\int_{-\infty}^{\infty} \frac{dz}{2\pi} e^{ix P^z z} \tilde{Q} (z, P^z,\eps),
\end{equation}
where the spacelike Wilson line is
\begin{equation}
W_z (z,0) = P \exp \left( -ig \int_0^z dz^{\prime} A^z (z^{\prime}) \right).
\end{equation}
The bilocal operator with a straight Wilson line is multiplicatively renormalizable. Accordingly, its matrix element 
$\tilde Q(z,P^z,\tilde\mu)$ may be renormalized in dimensional regularization or in another scheme $S$, such as 
the RI/MOM scheme, and can be written as
\begin{equation}
\tilde{Q}_B (z, P^z,\eps) = \tilde{Z}_S (z,\eps,\tilde{\mu})\, \tilde{Q}_R^S (z, P^z,\tilde{\mu}),
\end{equation}
where $\tilde{Z}_{S}$ is the counterterm in the scheme $S$.
After Fourier transformation, this product becomes a convolution for the quasi-PDF $\tilde q_S (x,P^z,\tilde{\mu})$.
 
For a nucleon moving with finite but large momentum
$P_z \gg \Lambda_{\text{QCD}}$, the quasi-PDF is matched onto the PDF as
\begin{equation} \label{matchingdef}
\tilde{q}_S (x,P^z,\tilde{\mu}) = \int_{-1}^1 \frac{dy}{|y|} C_S \left( \frac{x}{y}, \frac{\tilde{\mu}}{|y|P^z},
\frac{\mu}{|y|P^z} \right) q(y,\mu) +\mathcal{O}\left( \frac{M^2}{(P^z)^2}, 
\frac{\Lambda_{\mathrm{QCD}}^2}{(P^z)^2} \right).
\end{equation}
where $C_S$ is the matching coefficient, $M$ is the nucleon mass and the remaining terms in higher twists are suppressed. 
Note that the matching coefficient depends on the scheme $S$, while the PDF $q(y,\mu)$ is computed in the $\msbar$ scheme.
In Eq.~\eqref{matchingdef}, the dependence of the PDF on the factorization scale $\mu$ cancels between the
lightcone PDF and the matching coefficient on the right-hand side, up to higher-order corrections.
The appearance of the renormalization scale $\tilde\mu$ reflects the scheme used to renormalize the quasi-PDF.
 The PDF and the quasi-PDF share the same IR divergences, hence they are cancelled in matching. At the perturbative scales 
 $\mu$ and $\tilde{\mu}$, the matching coefficient $C_S$  can be computed to a desired order in $\as$.

We now turn to the partonic quasi-PDF in the RI/MOM scheme. We define the renormalization constant 
$\tilde{Z}_{\text{RI}} (z, p_R^z, \Lambda, \mu_R)$ by imposing a condition on the quasi-PDF using the massless 
quark state $|ps\rangle$  with momentum $p$ and spin $s$ with $p^2 \neq 0$. We consider the nonsinglet quark 
quasi-PDF, in which there is no mixing. The RI/MOM renormalization condition is 
\begin{align}\label{renocon}
&\tilde{Z}_{\text{RI}} (z, p_R^z, \Lambda, \mu_R)^{-1} \sum_s \langle ps| \overline{\psi} (z) \gamma^z W(z,0) \psi (0)
|ps\rangle \Big|_{p^2 = -\mu_R^2, p^z = p_R^z} \nonumber \\
& = \sum_s \langle ps| \overline{\psi} (z) \gamma^z W(z,0) \psi (0) |ps\rangle \Big|_{\text{tree}}
= 4p^z e^{-i z p^z} \zeta \Big|_{p^z = p_R^z},
\end{align}
where the partonic momentum is $p^{\mu} = (p^0, 0,0,p^z)$, the sum of the spinor over spin $s$ is
\begin{equation}
\zeta = \frac{1}{4p^z} \sum_s \overline{u}^s \gamma^z u^s,
\end{equation}
and $\mu_R$ is the renormalization scale employed in the RI/MOM scheme. The quantities in the ordinary RI/MOM scheme
are denoted by the subscript ``RI''. Here $\Lambda$ represents a collective 
representation for the UV cutoff. In lattice QCD, $\Lambda = a^{-1}$ is the inverse lattice spacing, and it is 
$\eps$ in dimensional regularization. The RI/MOM condition fixes the off-shellness $p^2=-\mu_R^2$ and the reference
momentum $p^z=p_R^z$.  When the resulting counterterm is applied to a quasi-PDF with external momentum $p^z$, 
the two longitudinal momenta $p^z$ and $p_R^z$ need not be equal.  That is, the ratio $\eta=p^z/p_R^z$ does not have 
to be equal to 1 in general. However, we first consider a special case $\eta =1$  because we can see the isolation of the 
scheme dependence clearly. After that, we extend the analysis for $\eta \neq 1$, 
which will be discussed in Section~\ref{etagen}.  

The factorized form for the quasi-PDF at the partonic level
in the RI/MOM scheme is
\begin{equation}
\tilde{q}_{\text{RI}} (x, p^z, p_R^z, \mu_R) = \int_{-1}^1 \frac{dy}{|y|} C_{\text{RI}} \left( \frac{x}{y}, 
\frac{\mu_R}{p_R^z}, \frac{\mu}{p^z} , \frac{p^z}{p_R^z}\right) q(y,\mu),
\end{equation}
where the hadronic momentum $P^z$ is equal to $yp^z$, and the suppressed higher-order terms are omitted.
There can be extra Dirac structure other than $\gamma^z$ because the matrix elements are evaluated at the 
off-shell value. To incorporate this, we use the external states with Dirac spinors $\overline{u} (p,s)$ and $u(p,s)$ and 
replace
\begin{equation}
\sum_s \overline{u} (p,s) \Gamma (z) u(p,s) \to \text{Tr} \left[ \fms{p} \Gamma (z) \right] \zeta,
\end{equation}
to define the off-shell prescription in evaluating the matrix element in Eq.~\eqref{renocon}.

We can compute the matching coefficient which relates the renormalized quasi-PDF in the ordinary RI/MOM scheme
to the renormalized PDF in the $\msbar$ scheme. The perturbative calculation in this paper is performed using dimensional 
regularization. Therefore power divergences of the nonlocal Wilson-line operator, such as the linear divergence proportional 
to the length of the gauge link that appears with a cutoff regulator, are absent from the continuum expressions considered here. 
With a cutoff regulator, this divergence is associated with the self-energy of the Wilson line and can be removed by a mass 
counterterm before the remaining logarithmic renormalization is carried out~\cite{Ji:2015jwa,Ji:2017oey,Ishikawa:2017faj}. 
The ordinary RI/MOM, minimal RI/MOM, and modified minimal RI/MOM schemes discussed below should therefore be 
understood as schemes for the bilocal operator after the power divergence has been removed.  If these schemes are applied 
to spatial correlations calculated in lattice QCD, the power divergence must be removed first in defining the renormalized 
bilocal operator, after which the conversion to the corresponding continuum scheme and the perturbative matching to the 
PDF may be performed~\cite{Constantinou:2017sej,Chen:2017mzz,Alexandrou:2017huk}. In this sense, the linear divergence 
is not part of the matching between the renormalized quasi-PDF and the lightcone PDF. It belongs to the renormalization of 
the bilocal operator from which the quasi-PDF is constructed. We do not consider this power subtraction further in this paper.

\begin{figure}[b]
\centering
\includegraphics[width=\columnwidth]{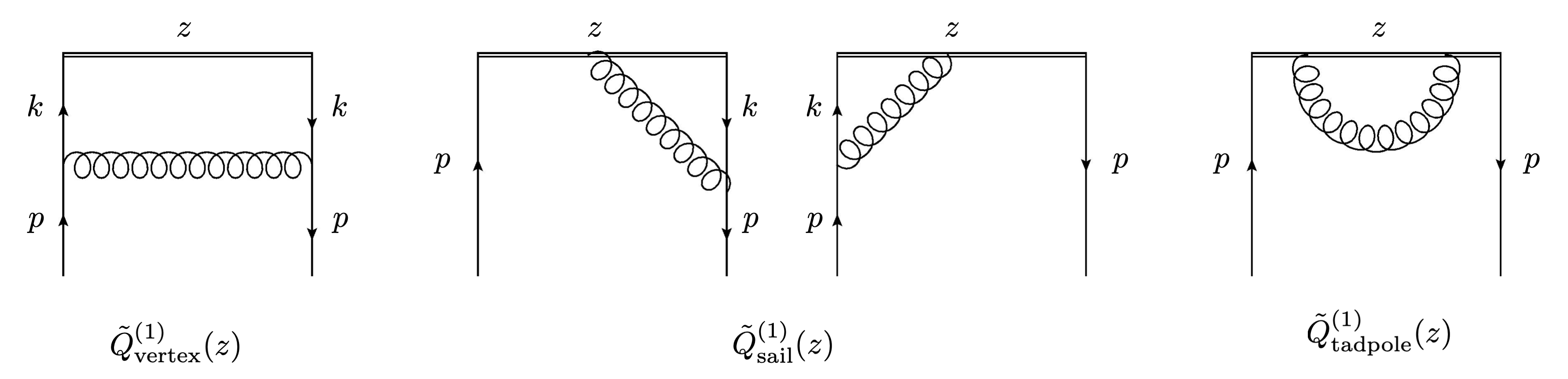}
\caption{ \label{fig1} 
Feynman diagrams for the quasi-PDF at one loop in coordinate space. The first one is the vertex diagram, the next two 
diagrams are the sail diagrams, and the last one is the tadpole diagram. The quark wave-function renormalization 
$\tilde{Q}^{(1)}_{\text{wf}} (z) $ is omitted.}
\end{figure}

Because the RI/MOM condition is imposed on off-shell matrix elements, the renormalized quasi-PDF and the counterterm 
depend on the gauge chosen in defining the scheme. This gauge dependence is compensated by that of the matching coefficient 
and therefore does not appear in the matched PDF. We use the Feynman gauge here. At tree-level, we have
\begin{equation}
\tilde{Q}^{(0)} (z, p^z) = 4p^z \zeta e^{-i zp^z},
\end{equation}
and the one-loop Feynman diagrams are shown in Fig.~\ref{fig1}.

 The full expressions for the one-loop diagrams in coordinate space are lengthy and are not needed for the argument of 
scheme separation developed here. They are given in Ref.~\cite{Stewart:2017tvs}, and they are independently
checked.  The sum of one-loop contributions $\tilde{Q}^{(1)} = \tilde{Q}^{(1)}_{\text{vertex}} + 
\tilde{Q}^{(1)}_{\text{sail}} +\tilde{Q}^{(1)}_{\text{tadpole}}  + \tilde{Q}^{(1)}_{\text{wf}}$ is written as~\cite{Stewart:2017tvs}
\begin{equation} \label{Qcor}
\tilde{Q}^{(1)} (z, p^z, -p^2) = \acz \int_{-\infty}^{\infty} dx \left( e^{-ix p^z z} - e^{-i p^z z}\right) h(x,\rho),
\end{equation}
where the function $h(x,\rho)$ is  
\begin{equation} \label{horig}
h(x,\rho) = \left\{ \begin{array}{ll}
\displaystyle \frac{1}{\sqrt{1-\rho}} \left[ \frac{1+x^2}{1-x} - \frac{\rho}{2 (1-x)} \right] \ln 
\frac{2x-1+\sqrt{1-\rho}}{2x-1-\sqrt{1-\rho}} -\frac{\rho}{4x(x-1) +\rho} +1, & x>1 \\
\displaystyle \frac{1}{\sqrt{1-\rho}} \left[  \frac{1+x^2}{1-x} - \frac{\rho}{2 (1-x)} \right] \ln 
\frac{1+\sqrt{1-\rho}}{1-\sqrt{1-\rho}} -\frac{2x}{1-x}, & 0<x<1, \\
\displaystyle \displaystyle \frac{1}{\sqrt{1-\rho}} \left[ \frac{1+x^2}{1-x} - \frac{\rho}{2 (1-x)} \right] \ln 
\frac{2x-1-\sqrt{1-\rho}}{2x-1+\sqrt{1-\rho}} +\frac{\rho}{4x(x-1) +\rho} -1, & x<0,
\end{array}
\right.
\end{equation}
Here we define
\begin{equation}
\rho \equiv \frac{-p^2 -i0}{p_z^2},
\end{equation}
which acts as an IR regulator, and the $-i0$ prescription in $\rho$ specifies the analytic continuation across the branch point. 

In the sum of the one-loop diagrams in Eq.~\eqref{Qcor}, there is no $1/\euv$ pole for finite $x$, hence  we can set 
$\eps=0$.  But when the behavior at $x=\pm\infty$ is considered, we have to be careful because UV poles may be developed.
Therefore when we construct the distributions at $\pm \infty$, we keep $\eps$, and it will be discussed in detail. As can be 
seen explicitly in Eq.~\eqref{Qcor}, the function $h(x,\rho)$ appears when the coordinate-space matrix element is written as a 
Fourier integral.  It should therefore not be identified by itself with the 
quasi-PDF. The two terms in the phase factor form a single combination under the Fourier transform and should not be treated separately. In momentum-fraction space, the second term generates the contribution localized 
at $x=1$, which combines with $h(x,\rho)$ to define the quasi-PDF with the appropriate prescription on distributions. The 
subtraction term proportional to $e^{-ip^z z}$ produces the $\delta(1-x)$ term after Fourier transformation, and this 
 turns $h(x,\rho)$ into the plus distribution on the full line. 
\begin{equation}
\tilde q_{\rm bare}^{(1)}(x,\rho)  = \acz\left[  h(x,\rho)  -  \delta(1-x)  \int_{-\infty}^{\infty}dy\,h(y,\rho) \right].
  \label{eq:qbare-before-plus}
\end{equation}
The relative coefficient of the two terms in Eq.~\eqref{eq:qbare-before-plus} is fixed by the explicit calculation of the 
coordinate-space matrix element in Eq.~\eqref{Qcor}. Consequently, the one-loop correction satisfies
\begin{equation}
 \int_{-\infty}^{\infty} dx\,  \tilde q_{\rm bare}^{(1)}(x,\rho)=0,
\end{equation}
which amounts to quark-number conservation at one loop.
 
 In the next section we rewrite 
Eq.~\eqref{eq:qbare-before-plus} using a plus distribution with respect to the point $x=1$. Thus the quasi-PDF 
in momentum space is not $h(x,\rho)$ itself, and this distinction is essential for the subsequent scheme 
decomposition. The parameter $\rho$ will be used below as an IR regulator for the quasi-PDF.  In constructing the 
matching coefficient one may either keep this IR regulator explicitly until it cancels against the lightcone PDF, or first 
pass to the corresponding on-shell expression after the common collinear structure has been separated. We will 
distinguish these two choices below. We first discuss the case $\eta =p^z/p_R^z =1$, where the 
result in the ordinary RI/MOM scheme takes the simple form $\tilde q_{\rm RI}^{(1)}(x,\rho,r_R)  =   \acz
\left[H(x,\rho)-H(x,r_R)\right]$, with $r_R=\mu_R^2/(p_R^z)^2$, where $H$ is a distribution to be specified. 
It makes the dependence on the reference momentum explicit and provides the
cleanest setting for defining the minimal and $\ribar$ schemes. For $\eta\neq1$, only the second term is modified 
by the rescaling $x-1\mapsto\eta(x-1)$. This deformation is a kinematic complication of the ordinary RI/MOM 
scheme, and it does  not invoke any new scheme prescription.

\section{Structure of the quasi-PDF in momentum-fraction space}\label{sep}
 
 By deriving Eq.~\eqref{eq:qbare-before-plus} explicitly, the bare quasi-PDF in $x$ space is written as the Fourier transform 
\begin{align} \label{bareqx}
\tilde{q}^{(1)} (x, p^z, \rho) & = \asc (4 p^z \zeta) \int_{-\infty}^{\infty} \frac{dz}{2\pi} e^{i x z p^z} 
\int_{-\infty}^{\infty} dy \left(e^{-iy p^z z} - e^{-i p^z z}\right) h(y,\rho) \nonumber \\
&= \asc (4\zeta) \int dy \left[ \delta (y-x) - \delta (1-x)\right] h(y,\rho) \nonumber \\
&= \asc (4\zeta)  \left[ h(x,\rho) -\delta (1-x) \int_{-\infty}^{\infty} dy h(y,\rho) \right]\equiv
\asc (4\zeta)  H(x,\rho),
\end{align}
where the function $h(x,\rho)$ in Eq.~\eqref{horig} enters the quasi-PDF as a 
distribution on the full line.  We define the distribution as
\begin{equation}  
H(x,\rho)\equiv [h(x,\rho)]_{+(1)}^{\mathbb R}.
\label{disful}
\end{equation}
The superscript specifies the domain on which the distribution is defined. More generally, for a function integrated over a 
domain $[a,b]$ that contains a singular point $x=c$, we define
\begin{equation}
\int_a^b dx\, [h(x,\rho)]_{+(c)}^{[a,b]} f(x) \equiv \int_a^b dx\, h(x,\rho)\left[f(x)-f(c)\right],
\label{eq:regional-plus}
\end{equation}
where $f(x)$ is a smooth test function. The subscript $+(c)$ indicates that the subtraction is made at the singular point $x=c$, 
while the superscript specifies the domain on which the prescription is defined.

For $H(x,\rho)$ in Eq.~\eqref{disful}, the singular point is $x=1$ and the domain is the full line. Thus 
$[\,\cdots\,]_{+(1)}^{\mathbb R}$ denotes the plus prescription on the full line centered at $x=1$. We will also use 
prescriptions defined separately in the regions $0<x<1$ and $x>1$. In particular,
\begin{equation}
\int_0^1 dx\, [h(x,\rho)]_{+(1)}^{[0,1]} f(x) = \int_0^1 dx\, h(x,\rho)\left[f(x)-f(1)\right],
\label{eq:regional-plus-01}
\end{equation}
and
\begin{equation}
\int_1^\infty dx\, [h(x,\rho)]_{+(1)}^{[1,\infty)} f(x) = \int_1^\infty dx\, h(x,\rho)\left[f(x)-f(1)\right].
\label{eq:regional-plus-1inf}
\end{equation}
For the region $x<0$, no plus prescription at $x=1$ is necessary. The notation for prescriptions at the boundaries 
$x=\pm\infty$ is distinct from that for the singular point at $x=1$. When a boundary prescription at $x=-\infty$ is needed, 
for example, we use $[\,\cdots\,]_{+(-\infty)}^{(-\infty,0]}$. The detailed definitions for these distributions are collected in
Appendix~\ref{app:dist-conv}.

With the plus prescription on the full line, the one-loop distribution satisfies
\begin{equation}
\int_{-\infty}^{\infty} dx\, H(x,\rho)=0,
\label{qcons}
\end{equation}
which states quark-number conservation. To see this, recall that the partonic analogue of the hadronic Fourier-transform 
relation in Eq.~\eqref{hadfour} is
\begin{equation}
\tilde q(x,p^z) = p^z\int_{-\infty}^{\infty}\frac{dz}{2\pi}\, e^{ixp^z z}\tilde Q(z,p^z).
\label{eq:partonic-Fourier}
\end{equation}
Therefore, the integral over $x$ yields
\begin{align}
\int_{-\infty}^{\infty} dx\, \tilde{q}(x,p^z) &= p^z \int_{-\infty}^{\infty} dx \int_{-\infty}^{\infty}\frac{dz}{2\pi}\,
e^{ixp^z z}\tilde{Q}(z,p^z) \nonumber \\
& = p^z\int_{-\infty}^{\infty} dz\, \delta(p^z z)\,\tilde{Q}(z,p^z) = \tilde{Q}(0,p^z).
\end{align}  
It states that the zeroth moment of the quasi-PDF projects the coordinate-space correlator onto the short-distance limit $z=0$. 
 At this point the Wilson line shrinks to unity, and the bilocal operator reduces to the local vector current. Therefore the 
 one-loop correction to the zeroth moment must vanish because the current is conserved. The two phase factors in 
 Eq.~\eqref{Qcor} combine under the Fourier
transform to give the plus prescription in Eq.~\eqref{disful}. The
subtraction term at $x=1$ is generated by the second phase factor, and
the resulting distribution has a vanishing zeroth moment. This is the
counterpart of the vanishing one-loop correction
to the local vector current at $z=0$ in the momentum-fraction space. After Fourier transformation to momentum-fraction space, 
the coefficient of the 
 collinear logarithm, the endpoint contribution associated with quark-number conservation, and the part associated with the 
 RI/MOM reference momentum can be discerned separately.

Now we identify the term, dependent on the reference momentum, that defines the ordinary RI/MOM counterterm. 
The counterterm in the ordinary RI/MOM scheme is determined by imposing the RI/MOM renormalization condition at 
the off-shell reference momentum. For $\eta=p^z/p_R^z=1$, the counterterm at one loop is
\begin{equation} 
\tilde{q}^{(1), \text{RI}}_{\text{CT}}(x,r_R) = -\acz H(x,r_R) = -\acz \left[h(x,r_R)\right]_{+(1)}^{\mathbb{R}},
\label{eq:qCT-RI}
\end{equation}
where $r_R=\mu_R^2/(p_R^z)^2$.  Adding this counterterm to the bare quasi-PDF gives the renormalized quasi-PDF
in the ordinary RI/MOM scheme,
\begin{equation} \label{offrenqri}
\tilde{q}^{(1)}_{\text{RI,off}} (x, \rho, r_R) = \acz \left[H(x,\rho)-H(x,r_R)\right].
\end{equation}
Here the subscript ``off'' indicates that the quasi-PDF is evaluated with an off-shell external quark state, with 
$\rho=-p^2/(p^z)^2$ kept finite as an IR regulator. This notation is different from the on-shell expression used below
in constructing the matching coefficient. The two terms in Eq.~\eqref{offrenqri} have different origins. The first term
is obtained from the matrix element evaluated with the external state and contains the collinear behavior relevant for 
matching. The second term is fixed by imposing the RI/MOM renormalization condition at the off-shell reference momentum. 
It is therefore through $H(x,r_R)$ that the RI/MOM reference parameter enters the renormalized quasi-PDF. The reference 
momentum is essential for the nonperturbative implementation of the RI/MOM prescription in lattice QCD, which we take as 
the starting point. The question addressed below is instead whether the function $H(x,r_R)$ must remain in the 
renormalized quasi-PDF, used subsequently for perturbative matching.

A finite separation of the two terms in Eq.~\eqref{offrenqri} would be arbitrary unless the part retained in the renormalized 
quasi-PDF is subject to the requirements mentioned above. First, it must contain the collinear logarithmic contribution 
required to cancel the corresponding infrared contribution in the lightcone PDF. Second, the finite-$x$, endpoint, and boundary 
terms must together define a distribution on the full line with a vanishing zeroth moment, so that quark number is
conserved. These conditions determine a common structure that must be retained, but they do not fix every finite term. 
The remaining freedom defines a finite choice of renormalization scheme. In the next section, we formulate this freedom in a 
general form and construct two explicit schemes.
  
 \section{Minimal and modified minimal RI/MOM schemes}
\label{minmom}

We now formulate a general class of finite RI/MOM-type schemes.  The ordinary RI/MOM prescription is retained as the 
nonperturbative renormalization condition for the spatial Wilson-line operator. The modification considered here is made 
afterward, at the perturbative matching stage. More precisely, we consider a finite transformation of the ordinary RI/MOM
quasi-PDF.  By finite, we mean that the difference between the two renormalized quasi-PDFs has a finite limit when the UV 
regulator is removed and therefore requires no additional UV subtraction. This does not mean that the difference is 
independent of the renormalization scales or regular as an ordinary function at $x=1$. It may contain logarithms involving the 
scales, and must be defined with the appropriate prescriptions at the endpoint and at infinity.

At one loop, let the quasi-PDF in a scheme $S$ be related to the ordinary RI/MOM quasi-PDF by
\begin{equation}
\tilde{q}_S^{(1)} = \tilde{q}_{\rm RI}^{(1)} - \Delta\tilde{q}_S .
\label{eq:finite-scheme-transformation}
\end{equation}
Since the bare quasi-PDF remains the same, the counterterm should be given by  
\begin{equation}
\tilde{q}_{\rm CT}^{S} = \tilde{q}_{\rm CT}^{\rm RI} + \Delta\tilde{q}_S,
\label{eq:finite-counterterm-transformation}
\end{equation}
which satisfies
\begin{equation}
\tilde{q}_{\rm bare}^{(1)} = \tilde{q}_{\rm CT}^{\rm RI} + \tilde{q}_{\rm RI}^{(1)} = \tilde{q}_{\rm CT}^{S} 
+ \tilde{q}_S^{(1)} .
\label{eq:bare-scheme-invariance}
\end{equation}
The matching coefficient undergoes the corresponding finite transformation so that the matched lightcone PDF is unchanged.
If a finite transformation involves a renormalization scale, it also changes the anomalous dimension.  At one loop, the derivative
with respect to $\mu_R$ yields
\begin{equation}
\frac{d}{d\ln\mu_R}\tilde{q}_S^{(1)} = \frac{d}{d\ln\mu_R}\tilde{q}_{\rm RI}^{(1)} -
\frac{d}{d\ln\mu_R}\Delta\tilde{q}_S .
\label{eq:finite-transformation-RG}
\end{equation}
Thus the $\mu_R$ dependence of the transformed quasi-PDF is not assigned independently.  It follows from the same 
transformation that relates the quasi-PDFs, counterterms, and matching coefficients.

Although a renormalization scheme is usually introduced by specifying its counterterm, it is more transparent here to 
proceed in the opposite direction. The modified schemes considered below are finite transformations of the ordinary 
RI/MOM scheme, so no additional ultraviolet divergence has to be removed. We therefore first specify the contribution to 
be retained in the renormalized quasi-PDF and then determine the corresponding finite counterterm from the difference. 
This order makes the criteria defining the modified schemes explicit.

We now determine which such transformations are admissible for quasi-PDF matching.  We specify a distribution 
$H_S(x,\rho)$ to be retained in the renormalized quasi-PDF,
\begin{equation}
\widetilde q_S^{(1)}(x,\rho)=\acz H_S(x,\rho).
\label{eq:qS-general}
\end{equation}
The retained distribution must contain the collinear logarithmic contribution in the physical region $0<x<1$, which is required 
to cancel the corresponding infrared contribution in the lightcone PDF. The one-loop quasi-PDF, however, has support on the 
entire real axis and contains logarithmic terms also in the regions $x>1$ and $x<0$. These terms are not fixed by collinear 
matching to the lightcone PDF. In defining the modified schemes, we choose to retain this logarithmic $x$ dependence 
outside the physical region rather than assign it to the counterterm. The remaining constants in the regions $x>1$ and $x<0$ 
are constrained  by the large-$\lvert x\rvert$ behavior of the logarithmic terms. This completion is not an additional 
physical requirement independent of the scheme choice. It follows from implementing the retained terms as a distribution on 
the full line.

These conditions and choices do not fix every finite term. We therefore write
 \begin{equation}
H_S(x,\rho)=H_{\min}(x,\rho)+F_S(x),
\label{eq:HS-general}
\end{equation}
where $H_{\min}$ contains the collinear logarithm in the physical region, the logarithmic terms chosen to be retained outside 
that region, and the completion required by their behavior at infinity. The distribution $F_S$ constitutes the finite terms 
that remain to be fixed by the choice of scheme, subject to the requirement
\begin{equation}
\int_{-\infty}^{\infty}dx\,F_S(x)=0.
\label{eq:FS-zero-moment}
\end{equation}

For $\eta=1$, the ordinary RI/MOM quasi-PDF is
\begin{equation}
\widetilde q_{\rm RI}^{(1)}(x,\rho,r_R) = \acz\left[H(x,\rho)-H(x,r_R)\right].
\end{equation}
Its difference from the quasi-PDF in scheme $S$ is therefore
\begin{equation}
\Delta\widetilde q_S(x,\rho,r_R) = \acz\left[H(x,\rho)-H(x,r_R)-H_S(x,\rho)\right].
\label{eq:DeltaqS-general}
\end{equation}
Using Eq.~\eqref{eq:finite-counterterm-transformation}, the counterterm in scheme $S$ is
\begin{align}
\tilde{q}_{\rm CT}^{S}(x,\rho) &= -\acz H(x,r_R)-\Delta\tilde{q}_S(x,\rho,r_R) \nonumber\\
&= \acz\left[-H(x,\rho)+H_S(x,\rho)\right].
\label{eq:CTS-general}
\end{align}
Since $H_S(x,\rho)$ contains the same collinear logarithmic contribution as $H(x,\rho)$, the logarithmic contribution cancels 
in their difference. The counterterm in Eq.~\eqref{eq:CTS-general} is therefore free of the collinear singularity retained in the 
renormalized quasi-PDF. Eq.~\eqref{eq:bare-scheme-invariance} then gives
\begin{equation}
\tilde{q}_S^{(1)}(x,\rho) = \acz H_S(x,\rho),
\end{equation}
in agreement with Eq.~\eqref{eq:qS-general}. Thus the dependence on the RI/MOM reference parameter cancels from the
renormalized quasi-PDF for any $H_S$ satisfying the conditions above. The collinear requirement ensures that the 
transformed counterterm is infrared finite, while the endpoint and boundary prescriptions ensure that the resulting 
expressions are well defined as distributions on the full line and maintain quark-number conservation. These requirements 
determine the common part $H_{\min}$ but leave the finite freedom represented by $F_S$. 
 
In summary, the bare quasi-PDF, which is independent of the RI/MOM reference momentum in the first place, is the sum 
of the renormalized quasi-PDF and the counterterm. In the ordinary RI/MOM scheme, each of these terms depends on the 
reference momentum, whereas the modified schemes make each term separately independent of it. Now we consider two 
choices. The minimal RI/MOM scheme corresponds to $F_{\rm mRI}=0$. The modified minimal RI/MOM scheme retains 
in addition a particular finite contribution in the physical region.

\subsection{Minimal RI/MOM scheme}

In the minimal RI/MOM scheme, we retain only the logarithmic splitting part in the physical region $0<x<1$, together with 
the logarithmic terms in the regions \(x>1\) and \(x<0\) and the constants required for their proper large-$\lvert x\rvert$ 
behavior. We first introduce the on-shell, or massless, $x$-space function used in the matching calculation. Following 
Ref.~\cite{Stewart:2017tvs}, we write this function as $h_0(x,\rho)$. The parameter $\rho$ is kept only in the collinear 
logarithm as an IR regulator.  Explicitly, it is 
\begin{equation} \label{h0}
h_0 (x,\rho) = \left\{ \begin{array}{ll}
\displaystyle \frac{1+x^2}{1-x} \ln \frac{x}{x-1} +1, & x>1, \\[0.7em]
\displaystyle  \frac{1+x^2}{1-x} \ln \frac{4}{\rho} -\frac{2x}{1-x}, & 0<x<1, \\[0.7em]
\displaystyle \frac{1+x^2}{1-x} \ln \frac{x-1}{x} -1, & x <0 .
\end{array}
\right.
\end{equation}
It is the on-shell function obtained from $h(x,\rho)$ by taking the limit $\rho\to 0$, with the endpoint singularity at $x=1$ to be 
treated by the plus prescription as discussed above. The off-shell function $h(x,\rho)$ in Eq.~\eqref{horig} is useful for 
displaying the RI/MOM subtraction, whereas $h_0(x,\rho)$ is the on-shell expression used after the 
common collinear infrared structure has been isolated.  This is not a change in  the RI/MOM renormalization condition, 
but a convenience for constructing the matching coefficient.  One could instead keep the off-shell function $h(x,\rho)$ in 
the renormalized quasi-PDF and carry the common infrared dependence through the calculation until
it cancels against the lightcone PDF.  The minimal choice, denoted by $h_{\mRI}(x,\rho)$, is obtained by retaining
the logarithmic collinear part in the physical region $0<x<1$.
\begin{equation} \label{hmri}
h_{\mRI}  (x,\rho) = \left\{ \begin{array}{ll}
\displaystyle \frac{1+x^2}{1-x} \ln \frac{x}{x-1}+1, & x>1, \\[0.7em]
\displaystyle  \frac{1+x^2}{1-x} \ln \frac{4}{\rho}, & 0<x<1, \\[0.7em]
\displaystyle \frac{1+x^2}{1-x} \ln \frac{x-1}{x}-1, & x <0 .
\end{array}
\right.
\end{equation}
In the regions $x>1$ and $x<0$, the constants in Eq.~\eqref{hmri} are kept because they cancel the constant large-$|x|$ 
limits of the logarithmic terms.  Without these constants, $h_{\mRI}(x,\rho)$ would approach a nonzero constant
as $x\to\pm\infty$, and the corresponding plus distribution on the full line would not have the desired behavior at infinity.
Thus the minimal prescription is minimal with respect to the physical collinear finite part, not with respect to the 
asymptotic part of the quasi-PDF.

We also introduce the corresponding distribution
\begin{equation}
H_{\mRI}(x,\rho) = [h_{\mRI}(x,\rho)]_{+(1)}^{\mathbb{R}},
\end{equation}
where the prescription of the plus distribution is the same as in Eq.~\eqref{disful}.   
The renormalized quasi-PDF $\tilde{q}^{(1)}_{\text{mRI}}$ in the minimal RI/MOM scheme at one loop is given by
\begin{equation} \label{mri-qpdf}
\tilde{q}^{(1)}_{\text{mRI}} (x,\rho) = \acz H_{\mRI} (x,\rho).
\end{equation}
In the physical region, this choice retains the collinear logarithmic contribution required to cancel the corresponding infrared
contribution in the lightcone PDF. The terms outside the physical region are retained to complete the distribution with the 
required large-$|x|$ behavior. In the general notation of Eq.~\eqref{eq:HS-general}, the minimal scheme corresponds to
$F_{\mRI}(x)=0$. Specializing the general result in Eq.~\eqref{eq:CTS-general} to $H_S=H_{\mRI}$ gives
\begin{equation}
\tilde{q}_{\rm CT}^{\rm mRI}(x,\rho) = \acz\left[-H(x,\rho)+H_{\mRI}(x,\rho)\right].
\label{eq:CT-mRI}
\end{equation}
The dependence on the RI/MOM reference parameter has already cancelled in the
general construction. The ordinary RI/MOM condition nevertheless remains the
nonperturbative renormalization condition from which this finite
transformation is made.
  
\subsection{Modified minimal RI/MOM scheme}

In the minimal RI/MOM scheme, we retain only the logarithmic splitting part in 
the renormalized quasi-PDF in the physical region. This prescription is useful because it isolates the part required for 
collinear matching, but it is not the only possible scheme for our purpose. For the next possible scheme, we retain the whole
$h_0 (x,\rho)$ with the finite terms that accompany the logarithmic part.  We call the resulting scheme the 
modified minimal RI/MOM scheme, or the $\ribar$ scheme. The minimal scheme keeps only the logarithmic structure, 
while the $\ribar$ scheme keeps a particular finite part useful for comparison with the standard quasi-PDF matching result.
In this respect, the distinction between the two modified schemes is analogous to that between the MS and $\msbar$ schemes. 
The minimal RI/MOM scheme retains only the part required by the defining prescription, whereas the $\ribar$ scheme also 
retains a specified finite contribution. The analogy is limited, however, since the finite difference between the two schemes 
here is a nontrivial function of $x$.

For the $\ribar$ scheme, we take $h_0(x,\rho)$ in Eq.~\eqref{h0} and define the corresponding distribution on the full 
line by
\begin{equation}
H_0(x,\rho)\equiv [h_0(x,\rho)]_{+(1)}^{\mathbb{R}} ,
\end{equation}
with the same prescription, as in Eq.~\eqref{disful}.  It satisfies
\begin{equation}
\int_{-\infty}^{\infty} dx\,H_0(x,\rho)=0,
\label{eq:H0-quark-number}
\end{equation}
which amounts to the one-loop statement of quark-number conservation.  Comparing $h_0(x,\rho)$ with $h_{\mRI}(x,\rho)$, 
we see that the two schemes are identical for the regions $x>1$ and $x<0$, and they differ only by 
\begin{equation}
h_0(x,\rho)-h_{\mRI}(x,\rho) = -\frac{2x}{1-x}\,\theta(x)\theta(1-x) .
\label{eq:h0-minus-hmin}
\end{equation}
Thus the difference between the minimal and $\ribar$ schemes is localized in
the physical region. In the notation of Eq.~\eqref{eq:HS-general}, the finite
choice defining the $\ribar$ scheme is
\begin{equation}
F_{\rbar}(x) = H_0(x,\rho)-H_{\mRI}(x,\rho) = \left[ -\frac{2x}{1-x}\theta(x)\theta(1-x) \right]_{+(1)}^{\mathbb R},
\label{eq:F-ribar}
\end{equation}
where we denote the quantities in the $\ribar$ scheme with $\rbar$.
It is independent of the collinear regulator and has a vanishing zeroth moment. The constants in the regions $x>1$ and $x<0$ 
are not part of this scheme difference, since they are already retained in $H_{\mRI}$ to maintain the large-$|x|$ behavior.
The renormalized quasi-PDF in the $\ribar$ scheme at one loop is therefore
\begin{equation}
\tilde{q}_{\rbar}^{(1)}(x,\rho) = \asc H_0(x,\rho).
\label{eq:q-ribar}
\end{equation} 
Specializing Eq.~\eqref{eq:CTS-general} to $H_S=H_0$ gives
\begin{equation}
\widetilde q_{\rm CT}^{\rbar}(x,\rho) = \acz\left[-H(x,\rho)+H_0(x,\rho)\right].
\label{eq:CT-ribar}
\end{equation}
As in the minimal scheme, the dependence on the RI/MOM reference momentum cancels as a consequence of the general 
finite transformation.

The $\ribar$ scheme therefore retains the plus distribution $H_0$ in the renormalized quasi-PDF. Compared with the minimal 
RI/MOM scheme, it includes the additional finite contribution $H_0-H_{\mRI}$, which changes both the quasi-PDF and the 
counterterm.  The defining feature of the $\ribar$ scheme is that it retains the conventional finite on-shell quasi-PDF 
contribution while removing the dependence on the RI/MOM reference momentum from the quantity used for matching. 
It differs from the minimal scheme by the finite distribution $F_{\rbar}=H_0-H_{\mRI}$.

\section{Matching coefficients, boundary terms, and RG behavior} \label{matco}

In this section we compute the matching coefficients in each scheme. In a scheme $S$, the matching relation between
the quasi-PDF $\tilde{q}_S$ and the PDF $q_{\msbar}$ is given by Eq.~\eqref{matchingdef}, and we expand each quantity
to order $\as$.  For convenience, we factor out $4\zeta$ for the 
external state, and expand the following quantities in powers of $\as$ as
\begin{align}
\tilde q_S(x) &= (4\zeta)\left[\delta(1-x)+\tilde{q}_S^{(1)}(x)+\cdots\right], \nonumber\\
q_{\overline{\rm MS}}(x,\mu) &= (4\zeta)\left[\delta(1-x)+q_{\overline{\rm MS}}^{(1)}(x,\mu)+\cdots\right].
\end{align}
The one-loop lightcone PDF in the $\msbar$ scheme is
\begin{equation}
q_{\msbar}^{(1)}(x,\mu) = \asc  \left[ \frac{1+x^2}{1-x}
\left(\ln\frac{\mu^2}{-p^2}-\ln[x(1-x)]\right) -(2-x) \right]_{+(1)}^{[0,1]} .
\label{eq:pdf-msbar}
\end{equation}
The matching coefficient is expanded as
\begin{equation}
C_S(x,\mu)=\delta(1-x)+C_S^{(1)}(x,\mu)+\cdots ,
\end{equation}
and the one-loop matching coefficient is obtained from
\begin{equation}  
C_S^{(1)}(x,\mu)=\tilde q_S^{(1)}(x)-q_{\msbar}^{(1)}(x,\mu).
\label{eq:CS-one-loop-definition}
\end{equation}

The matching coefficient is IR safe. In the preceding sections, the off-shell distribution $H(x,\rho)$ was used to describe the 
one-loop quasi-PDF, while $H(x,r_R)$ was used to keep the ordinary RI/MOM counterterm explicit. This formulation also 
made it possible to specify how finite terms are divided between the renormalized quasi-PDF and the counterterm. For the 
matching coefficient, however, the common collinear dependence of the quasi-PDF and the lightcone PDF must either be kept 
explicitly until it cancels or be represented by the corresponding on-shell expression. This is the point at which the connection 
with the matching construction enters, as in Ref.~\cite{Stewart:2017tvs}. The ordinary RI/MOM counterterm still contains 
the off-shell distribution $H(x,r_R)$, while the quasi-PDF side of the matching may be represented by $H_0(x,\rho)$ after the 
common infrared structure has been separated.  Thus there is no obstruction to using $H(x,\rho)$ in the renormalized 
quasi-PDF itself in the RI/MOM scheme.  Doing so simply requires one to carry the IR-regulated terms until their 
cancellation against the lightcone PDF is made explicit.  In the formulas below, this collinear structure is represented
by $H_{\mRI}(x,\rho)$ in the minimal scheme and by $H_0(x,\rho)$ in the $\ribar$ scheme.

\subsection{Matching coefficients}

In the minimal RI/MOM scheme the one-loop renormalized quasi-PDF is  given by
\begin{equation}
\tilde q_{\mRI}^{(1)}(x,\rho) = \asc\,H_{\mRI}(x,\rho)  = \asc
\begin{cases}
\displaystyle \frac{1+x^2}{1-x}\ln\frac{x-1}{x}-1, & x<0, \\[0.7em]
\displaystyle \left[ \frac{1+x^2}{1-x}\ln\frac{4}{\rho} \right]_{+(1)}^{[0,1]}, & 0<x<1, \\[0.7em]
\displaystyle \left[ \frac{1+x^2}{1-x}\ln\frac{x}{x-1}+1\right]_{+(1)}^{[1,\infty)}, & x>1 .
\end{cases}
\label{eq:q-minRI}
\end{equation}
Using Eq.~\eqref{eq:CS-one-loop-definition}, the corresponding matching coefficient is
\begin{equation}
C_{\mRI}^{(1)}(x) = \asc \begin{cases}
\displaystyle \frac{1+x^2}{1-x}\ln\frac{x-1}{x}-1, & x<0, \\[0.7em]
\displaystyle \left[ \frac{1+x^2}{1-x} \ln\frac{4x(1-x)(p^z)^2}{\mu^2} +(2-x) \right]_{+(1)}^{[0,1]},
& 0<x<1, \\[0.7em]
\displaystyle \left[\frac{1+x^2}{1-x}\ln\frac{x}{x-1}+1\right]_{+(1)}^{[1,\infty)}, & x>1 .
\end{cases}
\label{eq:CmRI}
\end{equation}
The dependence on the off-shell regulator $-p^2$ cancels in the physical region, as required for the matching coefficient.
The expressions above include only the prescription at the singular point $x=1$. They are therefore complete at finite $x$, 
but do not yet include the boundary prescriptions required at $x=\pm\infty$. In the next subsection, we extend the 
distributions to the whole line by including the prescriptions at the boundaries $x=\pm\infty$.

In the $\ribar$ scheme the renormalized quasi-PDF at one loop is instead chosen to be the massless on-shell expression,
\begin{equation}
\tilde q_{\rbar}^{(1)}(x,\rho) = \asc\,H_0(x,\rho)
  = \asc 
\begin{cases}
\displaystyle \frac{1+x^2}{1-x}\ln\frac{x-1}{x}-1, & x<0, \\[0.7em]
\displaystyle \left[ \frac{1+x^2}{1-x}\ln\frac{4}{\rho} -\frac{2x}{1-x} \right]_{+(1)}^{[0,1]}, & 0<x<1, \\[0.7em]
\displaystyle \left[\frac{1+x^2}{1-x}\ln\frac{x}{x-1}+1\right]_{+(1)}^{[1,\infty)}, & x>1 .
\end{cases}
\label{eq:q-ribar-explicit}
\end{equation}
The corresponding matching coefficient is therefore
\begin{equation}
C_{\rbar}^{(1)}(x) = \asc 
\begin{cases}
\displaystyle \frac{1+x^2}{1-x}\ln\frac{x-1}{x}-1, & x<0, \\[0.7em]
\displaystyle \left[ \frac{1+x^2}{1-x} \ln\frac{4x(1-x)(p^z)^2}{\mu^2} +(2-x)-\frac{2x}{1-x} \right]_{+(1)}^{[0,1]}, & 0<x<1,
\\[0.7em]
\displaystyle \left[\frac{1+x^2}{1-x}\ln\frac{x}{x-1}+1\right]_{+(1)}^{[1,\infty)}, & x>1 .
\end{cases}
\label{eq:Cribar}
\end{equation}
The term $(2-x)$ in the region $0<x<1$ originates from the $\msbar$ lightcone PDF, while the term $-2x/(1-x)$ belongs to the 
finite part retained in the $\ribar$ quasi-PDF.

The two matching coefficients differ by the same finite distribution that relates the two quasi-PDF schemes.
\begin{align}
C_{\rbar}^{(1)}(x)-C_{\mRI}^{(1)}(x) &= \tilde q_{\rbar}^{(1)}(x,\rho)-\tilde q_{\mRI}^{(1)}(x,\rho)
= \asc\left[H_0(x,\rho)-H_{\mRI}(x,\rho)\right] \nonumber\\
&= \asc\left[-\frac{2x}{1-x}\right]_{+(1)}^{[0,1]} .
\label{eq:scheme-difference}
\end{align}
Thus the difference between the two matching coefficients is entirely due to the finite term in the physical region, retained 
in the $\ribar$ quasi-PDF but not in the minimal quasi-PDF.  The constants for $x>1$ and $x<0$ do not contribute to the
scheme difference because they are already included in the minimal RI/MOM scheme.

\subsection{Completion at the boundaries of the full line}

We now complete the description of the quasi-PDF and matching coefficients by including the distributions at infinity. 
For finite $x$, the one-loop function $h(x,\rho)$ has no $1/\euv$ pole.  However, this does not mean that the quasi-PDF 
on the full line in $x$ is free of UV divergence.  In fact, the UV pole appears from the large-$|x|$ behavior. From the explicit 
form of $h(x,\rho)$, we find 
\begin{equation}
h(x,\rho) = -\frac{3}{2|x|} + \mathcal{O} \left(\frac{1}{x^2}\right), \qquad x\to \pm\infty.
\label{eq:h-large-positive}
\end{equation}
 Before the limit $\eps \to 0$ is taken, the corresponding large-$|x|$ behavior is given as
\begin{equation}
h(x,\rho;\eps ) \longrightarrow -\frac{3}{2}\frac{1}{|x|^{1+2\eps}}, \qquad x\to \pm\infty.
\label{eq:h-large-positive-eps}
\end{equation}
 Using the endpoint identities
\begin{align}
\frac{\theta(x-1)}{x^{1+2\eps}} &= -\frac{1}{2\euv} \frac{1}{x^2}\delta^+ \left(\frac{1}{x}\right)
+ \left(\frac{1}{x}\right)_{+(\infty)}^{[1,\infty)} +\mathcal{O}(\eps), \nonumber \\
\frac{\theta(-x)}{(1-x)^{1+2\eps}} &= -\frac{1}{2\euv} \frac{1}{(1-x)^2}\delta^+ \left(\frac{1}{1-x}\right)
+ \left(\frac{1}{1-x}\right)_{+(\infty)}^{(-\infty,0]} +\mathcal{O}(\eps),
\label{eq:plus-infinity-negative}
\end{align}
the plus distribution associated with $h(x,\rho;\eps)$ can be written as
\begin{align}
&H(x,\rho;\eps)  =   \theta(x-1) \left[ h(x,\rho)+\frac{3}{2x} \right] + \theta(x)\theta(1-x)\,h(x,\rho)
+ \theta(-x) \left[ h(x,\rho)+\frac{3}{2(1-x)} \right] \nonumber\\
& \hspace{0.7cm}
-\frac{3}{2} \left(\frac{1}{x}\right)_{+(\infty)}^{[1,\infty)} -\frac{3}{2}
\left(\frac{1}{1-x}\right)_{+(\infty)}^{(-\infty,0]} +\frac{3}{4\eps_{\rm UV}}
\left[ \frac{1}{x^2}\delta^+\!\left(\frac1x\right) + \frac{1}{(1-x)^2}\delta^+\!\left(\frac{1}{1-x}\right) \right].
\label{eq:H-full-eps}
\end{align}
Eq.~\eqref{eq:H-full-eps} displays only the additional structure of the distributions, associated with the large-$|x|$ 
behavior.  The singular terms at the endpoint $x=1$ are understood with the same plus prescription used in the 
definition of $H(x,\rho)$. That is, the first two terms in Eq.~\eqref{eq:H-full-eps} are to be interpreted 
as the plus distributions at $x=1$.  The combinations $h(x,\rho)+3/(2x)$ for $x>1$ and $h(x,\rho)+3/[2(1-x)]$ for $x<0$ 
then fall as $1/x^2$ at large $|x|$. The remaining terms explicitly represent the subtracted $1/|x|$ tail as
plus distributions at $+\infty$ and $-\infty$, together with the associated UV poles localized at the two boundaries.
The distribution $H(x,r_R;\epsilon)$ has the same boundary structure,
\begin{align}
&H(x,r_R;\eps) =  \theta(x-1) \left[ h(x,r_R)+\frac{3}{2x} \right] + \theta(x)\theta(1-x)\,h(x,r_R)
+ \theta(-x) \left[ h(x,r_R)+\frac{3}{2(1-x)} \right] \nonumber\\
&\hspace{0.7cm}
-\frac{3}{2} \left(\frac{1}{x}\right)_{+(\infty)}^{[1,\infty)} -\frac{3}{2} \left(\frac{1}{1-x}\right)_{+(\infty)}^{(-\infty,0]}
+\frac{3}{4\euv} \left[ \frac{1}{x^2}\delta^+\!\left(\frac1x\right) +
\frac{1}{(1-x)^2}\delta^+\!\left(\frac{1}{1-x}\right) \right] .
\label{eq:Hr-full-eps}
\end{align}
Therefore, for $\eta=p^z/p_R^z=1$, the renormalized quasi-PDF in the ordinary RI/MOM scheme is
\begin{align}
\tilde q_{\rm RI}^{(1)}(x,\rho,r_R) &= \acz\left[ H(x,\rho;\eps)-H(x,r_R;\eps) \right] \nonumber\\
&= \acz \left\{ \theta(x-1)\left[h(x,\rho)-h(x,r_R)\right] + \theta(x)\theta(1-x)\left[h(x,\rho)-h(x,r_R)\right]
\right. \nonumber\\ &\hspace{1.6cm}
\left.
+ \theta(-x)\left[h(x,\rho)-h(x,r_R)\right] \right\}_{+(1)}.
\label{eq:qRI-UV-cancel}
\end{align}
This result shows explicitly how the $1/\euv$ pole is canceled in the renormalized quasi-PDF in the ordinary RI/MOM
scheme on the whole line. The pole does not remain in the renormalized quasi-PDF, but the finite part of 
$H(x,r_R;\epsilon)$ still depends on $r_R=\mu_R^2/(p_R^z)^2$. We denote this finite part by $H_{\rm fin}(x,r_R)$. It is 
obtained from $H(x,r_R;\epsilon)$ by removing the UV-pole term localized at $x=\pm\infty$.
\begin{align}
H_{\rm fin}(x,r_R) &= \left\{ \theta(x-1)\left[h(x,r_R)+{3\over 2x}\right] +\theta(x)\theta(1-x)\,h(x,r_R)
+\theta(-x)\left[h(x,r_R)+{3\over 2(1-x)}\right] \right. \nonumber\\
&\hspace{1.0cm}\left. -{3\over2} \left({1\over x}\right)_{+(\infty)}^{[1,\infty)} -{3\over2}
\left({1\over 1-x}\right)_{+(\infty)}^{(-\infty,0]} \right\}_{+(1)} .
\label{eq:Hfin-explicit}
\end{align}
The plus prescription at $x=1$ is the same as in Eq.~\eqref{disful}.  The large-$|x|$ terms in Eq.~\eqref{eq:Hfin-explicit} 
are independent of $r_R$. All dependence on the RI/MOM reference momentum comes from $h(x,r_R)$.
Since the lightcone PDF has support only in the physical region, the quasi-PDF and the matching coefficient agree outside that 
region:
\begin{equation}
C_S^{(1)}(x)=\tilde q_S^{(1)}(x,\rho),
\qquad x<0\ \text{or}\ x>1,
\label{eq:C-q-outside}
\end{equation}
where $S=\mRI,\ribar$. Hence the same boundary prescriptions at $x=\pm\infty$ complete both quantities as distributions 
on the full line. Their explicit forms follow directly by replacing the finite-$x$ expressions in the preceding subsection with
 the corresponding completed distributions and need not be written separately.

The boundary terms are therefore not optional additions to the finite-$x$ expressions. They complete the quasi-PDF, 
counterterm, and matching coefficient as distributions on the full line and account for the UV pole that is invisible 
at finite $x$. Consequently, statements about quark-number conservation and renormalization-group (RG) behavior apply 
to the completed distributions rather than to the functions defined separately in the three finite-$x$ regions alone.

\subsection{Renormalization-group behavior}

We now derive explicitly the RG behavior implied by Eq.~\eqref{eq:finite-transformation-RG}.  
We first identify the scale dependence of the RI/MOM reference momentum in the ordinary scheme and then examine how 
it is transformed in the minimal and $\ribar$ schemes. At fixed $p_R^z$, the variables $\mu_R$ and 
$r_R=\mu_R^2/(p_R^z)^2$ are related by
\begin{equation}
\frac{d}{d\ln\mu_R}=2\frac{d}{d\ln r_R}.
\end{equation}
The one-loop anomalous dimension of the quasi-PDF in the ordinary RI/MOM scheme is therefore
\begin{equation}
\gamma_{\mu_R}^{\rm RI}(x,r_R) \equiv \frac{d}{d\ln\mu_R} \tilde q_{\rm RI}^{(1)}(x,\rho,r_R) =
-2\acz\,\frac{d}{d\ln r_R}H_{\rm fin}(x,r_R).
\label{eq:qRI-muR-dependence}
\end{equation}
The minus sign follows because $H_{\rm fin}(x,r_R)$ enters the renormalized quasi-PDF with a minus sign.
Evaluating the derivative separately in the three regions and writing the result in terms of inverse tangent functions for 
$r_R>1$ reproduces the piecewise function in Eq.~(D4) of Ref.~\cite{Stewart:2017tvs}. We do not repeat that expression here.  
 
In the minimal RI/MOM and $\ribar$ schemes, the renormalized quasi-PDFs are chosen to retain $H_{\mRI}(x,\rho)$ 
and $H_0(x,\rho)$, respectively.  These  functions do not contain $r_R$.  Therefore
\begin{equation}
\frac{d}{d\ln\mu_R}\tilde q_{\mRI}^{(1)}(x,\rho) = \frac{d}{d\ln\mu_R}\tilde q_{\rbar}^{(1)}(x,\rho) =0.
\label{eq:mRI-rbar-no-muR}
\end{equation}
This result follows directly from Eq.~\eqref{eq:finite-transformation-RG}. In both modified schemes, $\Delta\widetilde q_S$ 
contains the complete $r_R$-dependent part of the renormalized quasi-PDF in the ordinary RI/MOM scheme and therefore 
the renormalized quasi-PDFs in the modified schemes contain no $\mu_R$ dependence. The same finite transformation 
defines the corresponding counterterms and matching coefficients, which are likewise independent of $r_R$.

The relevant requirement is the consistency of the full matching relation in Eq.~\eqref{matchingdef}. Here we only track how 
the scale dependence is distributed between the renormalized quasi-PDF and the matching coefficient. Since the lightcone 
PDF does not depend on $\mu_S$, differentiation with respect to $\ln\mu_S$ gives
\begin{equation}
\frac{d}{d\ln\mu_S}\tilde q_S = \left[\frac{d}{d\ln\mu_S}C_S\right]\otimes q .
\label{eq:muS-RG-consistency}
\end{equation}
This equation is a consistency condition.  It does not require the $\mu_S$-dependence of each factor to be nonzero.  
In the ordinary RI/MOM scheme, $\mu_S=\mu_R$, and the quasi-PDF has the nontrivial anomalous dimension 
in Eq.~\eqref{eq:qRI-muR-dependence}.  The matching coefficient carries the corresponding $\mu_R$-dependence, so 
that Eq.~\eqref{eq:muS-RG-consistency} is satisfied.  In the minimal and modified minimal schemes, the renormalized
quasi-PDFs are independent of $r_R$, as shown in Eq.~\eqref{eq:mRI-rbar-no-muR}. The corresponding matching coefficients 
are related to the ordinary RI/MOM matching coefficient by the same finite transformation and therefore contain
no $r_R$-dependent contribution.  Hence both sides of 
Eq.~\eqref{eq:muS-RG-consistency} vanish for $\mu_S=\mu_R$.  

The dependence on the PDF factorization scale $\mu$ is different. The PDF is governed by the DGLAP evolution 
equation~\cite{Dokshitzer:1977sg,Gribov:1972ri,Altarelli:1977zs},
\begin{equation}
  {d\over d\ln\mu^2}q_{\msbar}^{(1)}(x,\mu)   =   \asc\,P_{qq}(x),
\end{equation}
where  
\begin{equation}
  P_{qq}(x)   =   \left[\frac{1+x^2}{1-x}\right]_{+(1)}^{[0,1]}   =   2\left(\frac{1}{1-x}\right)_{+}^{[0,1]}   -(1+x)
  +\frac{3}{2}\delta(1-x).
  \label{eq:Pqq-def}
\end{equation}
is the one-loop quark splitting distribution.  Since the quasi-PDF does not depend on $\mu$, the one-loop matching 
coefficient satisfies
\begin{equation}
{d\over d\ln\mu^2}C_S^{(1)}(x,\mu) = -\asc\,P_{qq}(x).
\end{equation}
This $\mu$-evolution is independent of how the terms associated with the RI/MOM reference momentum are 
distributed among the renormalized quasi-PDF, the counterterm, and the matching coefficient. Thus the minimal and modified 
minimal schemes do not change the RG content of the factorization formula. They differ only in the finite part retained in 
the renormalized quasi-PDF and, correspondingly, in the finite part included in the counterterm.  
For $\eta=p^z/p_R^z\neq1$, the function entering the ordinary RI/MOM counterterm is modified
because the external momentum and the RI/MOM reference momentum are different.  This general case is discussed 
in Sec.~\ref{etagen}.
 
\section{Scheme separation in coordinate space} \label{coint}
 
The finite transformation involved in defining the minimal and $\ribar$ schemes was formulated in momentum-fraction space.
We now examine its consequence in coordinate space, with particular attention to the coefficient of $\ln z^2$, which 
provides a diagnostic of where the short-distance logarithm appears in the matching relation. 
 
 \subsection{Quasi-PDFs in coordinate space}

Although the separation of the dependence on the reference momentum is achieved in momentum-fraction space, 
we now look back at the result in coordinate space in order to understand its effect on the spatial correlator and 
the matching coefficient. The purpose of this section is not to reproduce the full one-loop matrix element in coordinate 
space. Rather, we explain how the distributions introduced above enter the matrix element in coordinate space, and how 
the scheme dependence affects the renormalized quasi-PDF and the counterterm.

For the one-loop quasi-PDF $\tilde{q}^{(1)}(x)$, the corresponding matrix element in coordinate space is
\begin{equation}
\tilde{Q}^{(1)}(z,p^z,\tilde{\mu} ) =  \int_{-\infty}^{\infty} dx\, e^{-ixp^z z} \tilde{q}^{(1)}(x,p^z, \tilde{\mu}).
\label{eq:Q-from-q}
\end{equation}
If $\tilde{q}^{(1)}(x, p^z, \tilde{\mu})$ is expressed in terms of a plus distribution $[h(x)]_{+(1)}^{\mathbb{R}}$, $\tilde{Q}^{(1)}$ is 
proportional to
\begin{equation}
 \int_{-\infty}^{\infty} dx\, e^{-ixp^z z}[h(x)]_{+(1)}^{\mathbb{R}} =  \int_{-\infty}^{\infty} dx\,
\left(e^{-ixp^z z}-e^{-ip^z z}\right)h(x).
\label{eq:Q-plus}
\end{equation}
We can see that the second term in the Fourier transform is just the coordinate-space representation of the endpoint subtraction
in the plus prescription.  In particular, for any quasi-PDF in the form of the plus distributions used here,
\begin{equation}
\tilde Q^{(1)}(z=0,p^z, \tilde{\mu})=0,
\end{equation}
which is the coordinate-space form of the vanishing zeroth moment.  In the short-distance limit  $z=0$, or 
in terms of the Ioffe time $\nu=p^z z=0$,
the coefficient of the short-distance logarithm vanishes.  This is consistent
with quark-number conservation, since the bilocal operator reduces to the
local vector current. 

For $\eta=1$, the renormalized quasi-PDF in the ordinary RI/MOM scheme is the difference between the two distributions, 
$H(x,\rho)-H(x,r_R)$, as shown in Eq.~\eqref{offrenqri}.  Therefore the quasi-PDF in the ordinary RI/MOM scheme in 
coordinate space is
\begin{equation}
\tilde{Q}_{\rm RI}^{(1)}(z,p^z,\rho,r_R) = \asc (4p^z\zeta) \int_{-\infty}^{\infty} dx\,
\left(e^{-ixp^z z}-e^{-ip^z z}\right) \left[h(x,\rho)-h(x,r_R)\right].
\label{eq:coord-RI}
\end{equation}
There are two distinct operations in this expression.  The operation due to the RI/MOM renormalization condition  is the difference
$h(x,\rho)-h(x,r_R)$ between the physical and reference matrix elements.  On the other hand, the factor $e^{-ixp^z z}-e^{-ip^z z}$
implements the endpoint term from the plus prescription.  Therefore Eq.~\eqref{eq:coord-RI} corresponds to 
$H(x,\rho)-H(x,r_R)$ in momentum-fraction space, not merely to the function $h(x,\rho)-h(x,r_R)$.
When discussing the matching coefficient following Ref.~\cite{Stewart:2017tvs}, the quasi-PDF  may be represented by the 
on-shell function $h_0(x,\rho)$, while the RI/MOM subtraction still contains the off-shell function $h(x,r_R)$. The 
corresponding expression for the matching calculation in coordinate space is therefore obtained by replacing $h(x,\rho)$ by 
$h_0(x,\rho)$ in the first term.

In the minimal RI/MOM scheme, since the one-loop quasi-PDF is
\begin{equation}
\tilde q_{\mRI}^{(1)}(x,\rho) = \acz H_{\mRI}(x,\rho), \qquad
H_{\mRI}(x,\rho)=[h_{\mRI}(x,\rho)]_{+(1)}^{\mathbb{R}},
\label{eq:mRI-coordinate-space}
\end{equation}
in coordinate space, it is
\begin{equation}
\tilde Q_{\rm mRI}^{(1)}(z,p^z,\rho) = \asc (4p^z\zeta) \int_{-\infty}^{\infty} dx\,
\left(e^{-ixp^z z}-e^{-ip^z z}\right) h_{\mRI}(x,\rho).
\label{eq:coord-mRI}
\end{equation}
The difference between the results in the ordinary and the minimal RI/MOM schemes is
\begin{equation}
\Delta \tilde q_{\rm mRI}(x,\rho,r_R) = \acz \left[ H(x,\rho)-H(x,r_R)-H_{\mRI}(x,\rho) \right].
\label{eq:delta-mRI-coordinate-section}
\end{equation}
This part is included in the counterterm in the minimal RI/MOM scheme, as a result of which the resultant counterterm 
becomes also independent of $r_R$.

Similarly, the quasi-PDF in the $\ribar$ scheme in $x$ space is
\begin{equation}
\tilde q_{\overline{\rm RI}}^{(1)}(x,\rho) = \acz H_0(x,\rho), \qquad H_0(x,\rho)=[h_0(x,\rho)]_{+(1)}^{\mathbb{R}} .
\label{eq:barRI-coordinate-space}
\end{equation}
The corresponding quasi-PDF in coordinate space is
\begin{equation}
\tilde Q_{\overline{\rm RI}}^{(1)}(z,p^z,\rho)  = \asc (4p^z\zeta)  \int_{-\infty}^{\infty} dx\,
\left(e^{-ixp^z z}-e^{-ip^z z}\right) h_0(x,\rho).
\label{eq:coord-barRI}
\end{equation}
The remaining part relative to the ordinary RI/MOM result is
\begin{equation}
\Delta \tilde q_{\overline{\rm RI}}(x,\rho,r_R) = \acz \left[ H(x,\rho)-H(x,r_R)-H_0(x,\rho) \right],
\label{eq:delta-barRI-coordinate-section}
\end{equation}
and this part is included in the $\ribar$ counterterm.

Eqs.~\eqref{eq:coord-RI}, \eqref{eq:coord-mRI}, and \eqref{eq:coord-barRI} show the form of the scheme separations.  
In the ordinary RI/MOM scheme, the quasi-PDF in coordinate space contains the difference between the bare quasi-PDF and the  
term defined by the renormalization condition. In the minimal RI/MOM scheme, only $H_{\mRI}$ is retained in the 
renormalized quasi-PDF.  In the $\ribar$ scheme, $H_0$ is retained instead.   

\subsection{Coordinate-space interpretation of short-distance logarithms}
\label{subsec:coord-log}
 
Let us look into a specific topic, in which we consider the dependence on $\ln z^2$, and where it appears in each scheme. 
Schematically, the renormalized quasi-PDF in the ordinary RI/MOM scheme may be written as
\begin{equation}
 \tilde{Q}_{\rm RI}  =  \tilde{Q}_{\rm rem}  + \Delta \tilde{Q},
   \label{eq:coord-schematic-separation}
\end{equation}
where $\tilde{Q}_{\rm rem}$ denotes the part retained in the renormalized quasi-PDF in the modified scheme, while 
$\Delta\tilde{Q}$ denotes the part included in the corresponding counterterm. The coordinate-space representation provides 
a direct check of the scheme definition. It shows how the short-distance logarithm can be shifted between the factors in the 
matching relation by the finite transformation.

A useful way to examine this shift, motivated by coordinate-space renormalization prescriptions such as the 
hybrid scheme, is to look at the coefficient of the short-distance logarithm $\ln z^2$.  This logarithm is often used to 
diagnose how short-distance contributions are assigned in the spatial correlator.  Its interpretation, however, requires 
some care, because the short-distance limit and the lightcone limit with fixed Ioffe-time are different limits. 
 Let $\nu=p^z z$ be the Ioffe time~\cite{Radyushkin:2017cyf,Orginos:2017kos}. To identify the coefficient of $\ln z^2$, 
 we refer to the coordinate-space result obtained in dimensional regularization in Ref.~\cite{Izubuchi:2018srq}. In the 
 ordinary RI/MOM scheme, the UV pole is removed by the counterterm fixed through the renormalization condition. The 
 remaining IR-singular contribution contains
\begin{equation}
\frac{1}{\eir} \left(\frac{\mu^2 z^2 e^{2\gamma_E}}{4}\right)^{\eir} = \frac{1}{\eir} 
+ \ln\left(\frac{\mu^2 z^2 e^{2\gamma_E}}{4}\right) +\mathcal O(\eir).
\label{eq:IR-expand}
\end{equation}
Thus the infrared pole and the coordinate-space logarithm have the same coefficient. In the present calculation, the 
corresponding collinear singularity is regulated by $\rho$ and appears as $\ln(1/\rho)$ in momentum-fraction space. 
Its coefficient is $K(x)$, whose Fourier transform we denote by
\begin{equation}
\tilde K(\nu) = \int_0^1 dx\,e^{-ix\nu}K(x) = -\int_0^1 dx\,e^{-ix\nu}P_{qq}(x).
\label{eq:Ktilde-Ioffe}
\end{equation}
Up to the overall sign in our definition, $\tilde K(\nu)$ is the Fourier transform of the quark splitting function given in
Eqs.~(51) and (52) of Ref.~\cite{Izubuchi:2018srq}.

 In the short-distance limit one takes $z\to 0$, $p^z$ fixed,  and $\nu=p^z z\to0 $.
At this point the bilocal operator reduces to the local vector current, and quark-number conservation holds. As a result,
the coefficient of the short-distance logarithm vanishes.  This is equivalent to the statement that the coefficient function 
in coordinate space satisfies $\tilde{K}(0)=0$. This statement should be distinguished from the $z^2$ expansion with fixed 
Ioffe time. In that case one takes $z^2\to 0$, $\nu=p^z z$  fixed. The coefficient of $\ln z^2$ is therefore not evaluated at
$\nu=0$, but remains a nontrivial function of the Ioffe time, given by $\tilde K(\nu)$. The condition $\tilde{K}(0)=0$ follows 
from the normalization of the splitting function.  Thus there is no contradiction 
between the absence of $\ln z^2$ at short distance and the presence of a nonzero  $\ln z^2$ with fixed Ioffe time.
The explicit coordinate-space result of
Ref.~\cite{Izubuchi:2018srq} therefore gives
\begin{equation}
\left.\tilde{Q}^{(1)}(\nu,z^2)\right|_{\ln z^2} = \asc\,\tilde{K}(\nu) \ln\left(\frac{\mu^2 z^2 e^{2\gamma_E}}{4}\right).
\label{eq:Q-coordinate-log}
\end{equation}

In the ordinary RI/MOM scheme at $\eta=1$, the renormalized quasi-PDF is obtained by subtracting the counterterm 
determined from the matrix element evaluated at the RI/MOM reference momentum from the bare quasi-PDF. Since the 
coefficient $\tilde K(\nu)$ is independent of the infrared and reference parameters, it is the same for the matrix element 
evaluated at the external momentum and that evaluated at the RI/MOM reference momentum. The coordinate-space 
logarithm therefore cancels in the renormalized quasi-PDF.
\begin{equation}
\left.\tilde Q_{\rm RI}^{(1)}\right|_{\ln z^2} = \asc\left[\tilde K(\nu)-\tilde K(\nu)\right]
\ln\left(\frac{\mu^2 z^2 e^{2\gamma_E}}{4}\right) = 0.
\end{equation}
Thus, at fixed Ioffe time and $\eta=1$, the coefficient of $\ln z^2$ is absent from the renormalized quasi-PDF in the ordinary 
RI/MOM scheme and is retained in the RI/MOM counterterm. Consequently, the corresponding difference from the 
$\msbar$ PDF enters the matching coefficient.

In the modified schemes introduced in this work, the renormalized quasi-PDF is defined by the part retained after 
the dependence on the reference momentum is separated.  Both the minimal RI/MOM scheme and the $\ribar$ scheme retain 
the same collinear logarithmic part.  Therefore, when this retained part is viewed in coordinate space, the coefficient 
of the short-distance logarithm is
\begin{equation}
\tilde{K}_{\mRI}(\nu)  =  \tilde{K}_{\rbar}(\nu)  =  \tilde{K}(\nu).
\end{equation}
Equivalently,
\begin{equation}
  \left.  \tilde{Q}_{\mRI}^{(1)}  \right|_{\ln z^2} =  \left. \tilde{Q}_{\rbar}^{(1)} \right|_{\ln z^2}   =
\asc  \tilde{K}(\nu)   \ln{\mu^2 z^2 e^{2\gamma_E}\over4}.
\end{equation}
The difference between the minimal RI/MOM and $\ribar$ schemes is therefore not in the coefficient of $\ln z^2$, 
but in the finite terms retained in the renormalized quasi-PDF.  The counterterm is modified by the corresponding 
complementary contribution, so that the whole matching relation is unchanged.

The location of this logarithm has a direct implication for the matching coefficient.  In the ordinary RI/MOM scheme 
at $\eta=1$, the $\ln z^2$ term is absent from the renormalized quasi-PDF.  Therefore the difference between the 
ordinary RI/MOM quasi-PDF and the $\msbar$ lightcone PDF contains the full short-distance logarithm proportional to 
$\tilde{K}(\nu)$.  In this scheme the matching coefficient must therefore carry this logarithmic contribution.
By contrast, in the minimal RI/MOM and $\ribar$ schemes the retained quasi-PDF keeps the same $\ln z^2$ 
coefficient as the $\msbar$ lightcone PDF.  This logarithmic term then cancels between the quasi-PDF and 
the lightcone PDF in constructing the matching coefficient.  Thus the matching coefficients in these schemes do not 
contain this particular $\ln z^2$ mismatch.  Their difference is instead controlled by the choice of the retained finite distribution.

This is the practical advantage of the minimal and modified minimal organizations. They keep the universal collinear 
logarithmic structure in the quasi-PDF used for matching, so that the matching coefficient is not burdened with the 
logarithm that is already present in the lightcone PDF.  The ordinary RI/MOM scheme is, of course, a valid scheme, but 
its cancellation of the $\ln z^2$ term inside the renormalized quasi-PDF transfers the corresponding short-distance
logarithm to the perturbative matching coefficient.

 \section{General case $p^z \neq p_R^z$} \label{etagen}

We have formulated the modified schemes for $\eta=1$, where the separation between the term evaluated with the external 
momentum and the term fixed by the RI/MOM renormalization condition is most transparent. We now examine whether the 
construction remains valid when the external longitudinal momentum differs from the RI/MOM reference momentum. In this 
case, the function entering the ordinary RI/MOM counterterm is deformed, but the conditions defining the retained distribution 
are unchanged.

For $\eta=1$, the renormalized quasi-PDF in the RI/MOM scheme takes the simple form $\tilde{q}_{\rm RI}(x,\rho,r_R)
= \tilde{q}_{\rm bare}(x,\rho)- \tilde{q}_{\rm bare}(x,r_R)$. So the dependence on the reference momentum is isolated in 
the second term $\tilde{q}_{\rm bare}(x,r_R)$, and it takes the simple form in the ordinary RI/MOM scheme as
\begin{equation}
\tilde{q}_{\rm RI}^{(1)}(x,\rho,r_R)   = \acz \left[ H(x,\rho)-H(x,r_R) \right],
\end{equation}
where $H(x,a) = [h(x,a)]_{+(1)}^{\mathbb{R}}$ is the plus distribution associated with the function $h(x,a)$. We now explain 
how this expression is modified when $p^z$ and $p_R^z$ are different.

The conceptual separation introduced in the previous sections is unchanged.  In particular, the ordinary RI/MOM result is still 
decomposed into a renormalized part and a counterterm.  What changes is the form of the 
function entering the ordinary RI/MOM counterterm.  The RI/MOM renormalization condition is imposed at the 
momentum $p_R^z$, whereas the quasi-PDF is Fourier transformed with respect to the physical momentum $p^z$. 

For $\eta\neq 1$, the origin of the shifted argument is most transparent in coordinate space. After factoring out the common
prefactor $(4\zeta)$, the one-loop contribution to the on-shell quasi-PDF can be written as
\begin{equation}
\int_{-\infty}^{\infty} dx\,
\left(e^{-ixp^z z}-e^{-ip^z z}\right) h_0(x,\rho) .
\end{equation}
The second term subtracts the contribution at the partonic endpoint $x=1$, and the phase appearing in the loop correction can 
be written as
\begin{equation}
e^{-ixp^z z}-e^{-ip^z z} = e^{-ip^z z}\left[e^{-i(x-1)p^z z}-1\right].
\end{equation}
The overall factor $e^{-ip^z z}$ is fixed by the external momentum of the quasi-PDF. In the term evaluated at the RI/MOM 
momentum $p_R^z$, only the phase measured relative to the endpoint is expressed in terms of $p_R^z$. We therefore define 
$x_R$ through
\begin{equation} \label{pzpr}
e^{-i(x-1)p^z z} = e^{-i(x_R-1)p_R^z z}.
\end{equation}
With this definition,
\begin{equation}
e^{-ip^z z}\left[e^{-i(x_R-1)p_R^z z}-1\right] = e^{-ip^z z}e^{ip_R^z z}
\left(e^{-ix_Rp_R^z z}-e^{-ip_R^z z}\right).
\end{equation}
The momentum $p_R^z$ thus enters through the phase measured from the endpoint, while the overall phase remains 
determined by the external momentum $p^z$.

This shows that $p_R^z$ enters only through the phase measured from $x=1$, while the overall endpoint phase remains tied 
to the external momentum $p^z$. From Eq.~\eqref{pzpr}, we have 
\begin{equation}
 (x_R-1)p_R^z=(x-1)p^z .
\end{equation}
With $\eta=p^z/p_R^z$, we obtain
\begin{equation}
  x_R = 1+\eta(x-1),   \qquad   dx_R=|\eta|\,dx .
\end{equation}
Note that the point $x=1$ is fixed by this transformation.  Thus the prescription for the plus distribution remains centered 
at $x=1$, but the function entering the ordinary RI/MOM counterterm is stretched or compressed around this point. 
Including the Jacobian, we define
\begin{equation}
H(x,r_R)^{(\eta)}  \equiv  \left[   |\eta|\,h\bigl(1+\eta(x-1),r_R\bigr)   \right]_{+(1)}^{\mathbb{R}},
\label{eq:deformed-H-def}
\end{equation}
and its action on a test function $f(x)$ is
\begin{align}
  \int_{-\infty}^{\infty} dx\,   H(x,r_R)^{(\eta)} f(x)   ={}&   \int_{-\infty}^{\infty} dx\,   |\eta|\,h\bigl(1
  +\eta(x-1),r_R\bigr) \bigl[f(x)-f(1)\bigr] .
  \label{eq:deformed-H-test}
\end{align}
For positive $\eta$, using the change of variables $y=1+\eta(x-1)$, we obtain the useful form
\begin{equation}
\int_{-\infty}^{\infty} dx\,   H(x,r_R)^{(\eta)} f(x)   =   \int_{-\infty}^{\infty} dy\,   h(y,r_R)   \left[
f\left(1+\frac{y-1}{\eta}\right)-f(1)   \right] .
\label{eq:deformed-H-change-variable}
\end{equation}
This form shows precisely where the effect of $\eta\neq 1$ enters. The function $h(y,r_R)$ entering the ordinary RI/MOM 
counterterm is unchanged and is expressed in its own momentum-fraction variable $y$, while the test function is evaluated at 
the rescaled point $x=1+(y-1)/\eta$.  Thus the point $x=1$ is fixed, so the prescription for the plus distribution remains centered 
at $x=1$, while distances from this point are stretched or compressed by the factor $\eta$.  Put it in a different way, the function 
$H(x,\rho)$ associated with the external momentum is not deformed. Only the function entering the ordinary RI/MOM 
counterterm is changed by the mismatch between the external momentum $p^z$ and the reference momentum $p_R^z$.

With this notation, the renormalized quasi-PDF in the ordinary RI/MOM scheme for general $\eta$ at one loop is
\begin{equation} \label{genetari}
\tilde{q}_{\rm RI}^{(1)}(x,\rho,r_R,\eta)   = \acz \left[ H(x,\rho)-H(x,r_R)^{(\eta)} \right].
\end{equation}
This is the direct generalization of the result with $\eta=1$. When $\eta$ is set to one, $H(x,r_R)^{(\eta)}$ reduces to $H(x,r_R)$, 
and Eq.~\eqref{genetari} becomes the $\eta=1$ result proportional to $H(x,\rho)-H(x,r_R)$. For $\eta\neq1$, the difference from 
the $\eta=1$ case is entirely contained in $H(x,r_R)^{(\eta)}$. Before the plus prescription is imposed, $h(x,r_R)$ is replaced by 
$|\eta|h(1+\eta(x-1),r_R)$. This deformation changes the locations of the boundaries separating the three regions.  For $\eta>0$, 
the regions of the variable $x_R$ map as
\begin{equation}
 x_R<0     \Longleftrightarrow x<1-\frac1\eta,\ \ 0<x_R<1    \Longleftrightarrow 1-\frac1\eta<x<1,\ \
x_R>1  \Longleftrightarrow x>1 .
 \label{eq:eta-region-map}
\end{equation}
Therefore the general expression is more cumbersome if it is written out in each region.  This complication is purely 
kinematic.  It does not alter the logic of the scheme separation.

We now apply the same definitions of the minimal and modified minimal schemes to the deformed ordinary 
RI/MOM result.  In the minimal RI/MOM scheme, the difference is
\begin{equation}
\Delta \tilde{q}_{\rm mRI}(x,\rho,r_R,\eta)   = \acz \left[  H(x,\rho)-H(x,r_R)^{(\eta)}-H_{\mRI}(x,\rho)\right].
  \label{eq:delta-min-general-eta}
\end{equation}
The ordinary RI/MOM counterterm is now
\begin{equation}
\tilde{q}_{{\rm CT}}^{\rm RI}(x,r_R,\eta)   = -\acz  H(x,r_R)^{(\eta)} .
  \label{eq:CT-RI-general-eta}
\end{equation}
The counterterm in the minimal RI/MOM scheme is defined, as before, by absorbing the finite difference into the ordinary 
counterterm as
\begin{equation}
\tilde{q}_{{\rm CT}}^{\rm mRI}(x,\rho,r_R,\eta)   =   \tilde{q}_{{\rm CT}}^{\rm RI}(x,r_R,\eta)
- \Delta \tilde{q}_{\rm mRI}(x,\rho,r_R,\eta).
  \label{eq:CT-min-general-eta-def}
\end{equation}
Using Eqs.~\eqref{eq:delta-min-general-eta} and \eqref{eq:CT-RI-general-eta}, we obtain
\begin{equation}
\tilde{q}_{{\rm CT}}^{\rm mRI}(x,\rho)   =\acz \left[ -H(x,\rho)+H_{\mRI}(x,\rho)\right],
  \label{eq:CT-min-general-eta-result}
\end{equation}
so that the renormalized quasi-PDF is
\begin{equation}
\tilde{q}_{\mRI}^{(1)} (x,\rho) =\acz  H(x,\rho)+\tilde{q}_{{\rm CT}}^{\rm mRI}(x,\rho)   = \acz H_{\mRI}(x,\rho).
\end{equation}
Thus the deformation changes the ordinary RI/MOM result and the difference assigned to the counterterm, but not
the final quasi-PDF in the minimal RI/MOM scheme. 

The $\ribar$ scheme is treated in exactly the same way.  The difference is
\begin{equation}
  \Delta \tilde{q}_{\overline{\rm RI}}(x,\rho,r_R,\eta)   =   \acz\left[H(x,\rho)-H(x,r_R)^{(\eta)}-H_0(x,\rho)\right].
  \label{eq:delta-bar-general-eta}
\end{equation}
where $H_0(x,\rho)$ is the plus distribution retained in the modified minimal scheme. The corresponding counterterm is
\begin{equation}
\tilde{q}_{{\rm CT}}^{\overline{\rm RI}}(x,\rho,r_R,\eta)    =   \tilde{q}_{{\rm CT}}^{\rm RI}(x,r_R,\eta)
  - \Delta \tilde{q}_{\overline{\rm RI}}(x,\rho,r_R,\eta)    = \acz \left[-H(x,\rho)+H_0(x,\rho)\right],
  \label{eq:CT-bar-general-eta-result}
\end{equation}
yielding
\begin{equation}
 \tilde{q}_{\rbar}^{(1)} (x,\rho) =\acz H(x,\rho)+\tilde{q}_{{\rm CT}}^{\overline{\rm RI}}(x,\rho)   =\acz  H_0(x,\rho).
\end{equation}

These results show that the cancellation of the reference parameters is not a special consequence of setting $p^z=p_R^z$. 
For $\eta\neq1$, the ordinary RI/MOM contribution depends both on the off-shell reference parameter $r_R$ and on the 
deformation generated by $\eta=p^z/p_R^z$. After the finite scheme transformation is applied, the dependence remains 
neither in the counterterms, nor in the renormalized quasi-PDFs of the minimal and modified minimal schemes. The 
construction is therefore independent of the particular reference kinematics used to express the ordinary RI/MOM result. 
The ratio $\eta$ does not require an additional renormalization scheme. It changes the part of the ordinary RI/MOM result 
evaluated at the reference momentum, and therefore the finite difference included in the counterterm, but leaves unchanged 
the distributions $H_{\mRI}$ and $H_0$ retained in the two modified schemes.

\section{Relation to the hybrid-ratio prescription}
\label{sec:hybrid-comparison}

A related method of removing the dependence on the RI/MOM reference parameters was considered in 
Ref.~\cite{Chou:2022drv} in the hybrid-RI/MOM scheme. In this prescription, RI/MOM renormalization is applied for spatial 
separations $|z|<z_s$, while a mass subtraction is used for $|z|>z_s$. It was shown that, when the longitudinal component 
of the RI/MOM reference momentum is set to zero $p_R^z=0$, and the short-distance condition $\mu_R z_s\ll 1$
is imposed, the hybrid-RI/MOM prescription reduces to the hybrid-ratio prescription, apart from the charge associated 
with the corresponding PDF.

The mechanism responsible for this result is most transparent in coordinate space. The matrix element entering the RI/MOM 
renormalization condition depends on the Ioffe time associated with the reference momentum $\nu_R=p_R^z z$.
For $p_R^z=0$, this variable vanishes for every $z$. The matrix element entering the RI/MOM renormalization condition 
therefore has no nontrivial dependence on the longitudinal momentum flowing along the Wilson line. In addition, when 
$\mu_R z\ll1$, the dependence on the off-shell reference 
scale is suppressed by powers of $\mu_R^2z^2$. The short-distance RI/MOM renormalization factor then becomes proportional to 
the zero-Ioffe-time matrix element used in the ratio scheme. Consequently, the dependence on $p_R^z$ and $\mu_R$ disappears 
from the corresponding hybrid matching coefficient.

This result should not be interpreted as the absence of all scheme dependence. The matching coefficient obtained in this limit is 
that of the hybrid-ratio scheme and retains the dependence on the separation scale $z_s$ that specifies the boundary between 
the two parts of the hybrid prescription. What disappears is the dependence associated with the RI/MOM reference momentum 
and virtuality. The choice $p_R^z=0$ therefore replaces the ordinary RI/MOM renormalization condition at finite reference 
Ioffe time by a prescription at zero reference Ioffe time, rather than demonstrating that the ordinary RI/MOM matching 
coefficient is independent of its reference parameters. 

The prescription obtained by setting $p_R^z=0$ is not a regular limit of the momentum-fraction formulation developed in 
Sec.~\ref{etagen}. The variables used there behave as
\begin{equation}
r_R=\frac{\mu_R^2}{(p_R^z)^2}\longrightarrow\infty, \qquad \eta=\frac{p^z}{p_R^z}\longrightarrow\infty
\end{equation}
when $p_R^z\to0$. Correspondingly, the transformation $x_R=1+\eta(x-1)$
has no finite limit for a generic value of $x$. The prescription with $p_R^z=0$ must therefore be defined directly in coordinate 
space and cannot be obtained by substituting a limiting value of $r_R$ or $\eta$ into the ordinary momentum-fraction RI/MOM 
result.

The construction in the present work addresses a different question. We retain a finite longitudinal reference momentum and 
begin with the quasi-PDF in the ordinary RI/MOM scheme, whose dependence on $r_R$ is nontrivial and, for $p^z\neq p_R^z$, 
is accompanied by the deformation parameter $\eta$. We then identify the part associated with the RI/MOM reference momentum 
and include it in the counterterm through a finite modification of the renormalization prescription. Thus the dependence on the 
RI/MOM reference momentum is not removed by a special choice of kinematics, but is separated explicitly at general finite 
reference momentum.

The two approaches may nevertheless lead to similar matching coefficients because they remove the dependence on the 
RI/MOM reference momentum in different ways. In the hybrid construction with $p_R^z=0$, the dependence on the reference 
Ioffe time is absent from the outset. In the minimal and modified minimal schemes introduced here, the part associated with 
the RI/MOM reference momentum is separated from the renormalized quasi-PDF at finite $p_R^z$.
The resemblance of the remaining terms is therefore consistent 
with their interpretation as the part of the matching relation that is independent of the RI/MOM reference momentum. The 
present construction further shows how this part emerges from the ordinary RI/MOM result without restricting the longitudinal 
reference momentum to zero.

\section{Conclusions}
\label{conc}

In this paper we have formulated finite transformations of the ordinary RI/MOM scheme for quasi-PDFs at one loop. In the 
ordinary prescription, the off-shell reference momentum enters the renormalized quasi-PDF through the term fixed by the 
RI/MOM renormalization condition. We retain this condition as the nonperturbative renormalization prescription, but separate 
that term from the distribution subsequently used for perturbative matching.

 The retained distribution in a scheme $S$ was written as $H_S(x,\rho)=H_{\rm min}(x,\rho)+F_S(x)$, 
where $H_{\rm min}$ contains the common collinear structure and the terms required for its completion as a distribution on 
the full line. The finite distribution $F_S$ represents the remaining scheme freedom and has a vanishing zeroth moment. Once 
$H_S$ is specified, the renormalized quasi-PDF and counterterm are fixed as
\begin{equation}
\tilde{q}_S^{(1)}(x,\rho) =\acz H_S(x,\rho),\ \ \tilde{q}_{\rm CT}^{S}(x,\rho)
=\acz\left[-H(x,\rho)+H_S(x,\rho)\right].
\end{equation}
The minimal RI/MOM scheme corresponds to $F_{\mRI}=0$, while the $\ribar$ scheme retains the additional finite 
contribution $H_0-H_{\rm min}$ in the physical region.
The corresponding matching coefficients differ by the same finite distribution $H_0-H_{\rm min}$. Since this difference 
contains no collinear logarithm, the two schemes have the same logarithmic structure and the same one-loop anomalous 
behavior. The finite choice changes the quasi-PDF, counterterm, and matching coefficient separately, but as a whole, they leave 
the matched lightcone PDF unchanged.

The analysis in coordinate space provides an independent consequence of the scheme construction. In the ordinary RI/MOM 
scheme at $\eta=1$, the coefficient of the short-distance logarithm $\ln z^2$ cancels between the matrix elements evaluated at 
$p^z$ and $p_R^z$, and is therefore absent in the renormalized quasi-PDF. In the minimal and modified minimal schemes, the
common logarithmic contribution is retained in the quasi-PDF, and it is absent in the matching coefficient.

The expressions for finite $x$ must be completed by the contributions localized at $x=\pm\infty$. Although the one-loop 
matrix elements have a smooth $\eps\to0$ limit at finite $x$, the large-$|x|$ tails contain ultraviolet terms that become visible 
only after they are expanded before the limit is taken. These boundary terms complete the quasi-PDFs and counterterms as 
distributions on the full line and are required for quark-number conservation and renormalization-group behavior.

The construction of the modified schemes remains valid for $\eta=p^z/p_R^z\neq1$. In this case the function entering the 
ordinary RI/MOM counterterm is deformed according to $x-1\mapsto\eta(x-1)$ with the corresponding Jacobian, but the 
retained distributions $H_{\mRI}$ and 
$H_0$ are unchanged. The parameters $r_R$ and $\eta$ therefore cancel from the counterterms and renormalized quasi-PDFs
of the two modified schemes. This cancellation is not a special consequence of the choice $\eta=1$, and the general choice 
$\eta\neq1$ does not call for an additional renormalization scheme.

The scheme separation discussed here is formulated in dimensional regularization and therefore concerns the perturbative
finite parts of the continuum RI/MOM prescription.  In a cutoff implementation, such as a lattice
calculation, the bilocal operator contains a power divergence.  That power divergence must be subtracted from the lattice 
spatial correlator before the conversion to any of the continuum RI/MOM-type schemes is performed. But this is 
a separate issue from the RI/MOM scheme dependence analyzed in this work.

The comparison with the hybrid-RI/MOM prescription provides another perspective on the dependence on the RI/MOM 
reference momentum. When $p_R^z=0$ and $\mu_Rz_s\ll1$, the hybrid-RI/MOM prescription becomes equivalent to the 
hybrid-ratio prescription, apart from the PDF charge, and the dependence on the RI/MOM reference parameters disappears. 
This is achieved by choosing zero reference Ioffe time. By contrast, the construction presented here retains finite $p_R^z$ and 
separates the part associated with the RI/MOM reference momentum within the ordinary RI/MOM result. The two approaches 
therefore eliminate the dependence on $p_R^z$ in different ways.  A quantitative comparison of the resulting finite matching
coefficients, including the dependence on the hybrid separation scale $z_s$, would be useful. 

We conclude that the dependence of the ordinary RI/MOM quasi-PDF on the reference momentum can be separated from 
the part retained for perturbative matching.  Subject to the requirements from collinear matching, quark-number conservation, 
and the definition of distributions on the full line, the finite transformation removes this dependence from the renormalized 
quasi-PDF, the counterterm, and the matching coefficient in the modified schemes. The minimal and modified minimal 
schemes provide two explicit realizations of this construction. Their counterterms, matching coefficients, boundary 
contributions, and coordinate-space logarithms follow consistently from the chosen retained distribution.

 \appendix

\section{Distribution conventions on the full line}
\label{app:dist-conv}

We summarize the conventions of the distributions used in the main text. Since quasi-PDFs have support on the full real line, 
three distinct issues must be distinguished. They are the plus prescription at the endpoint $x=1$, its implementation in 
different regions of $x$, and the boundary prescriptions at $x=\pm\infty$.

\subsection{Plus distributions at $x=1$}

For a function $f(x)$ with an integrable singularity away from $x=1$ and an endpoint singularity at $x=1$, we define 
the plus distribution on the full line by
\begin{equation}
\int_{-\infty}^{\infty} dx\, \left[f(x)\right]_{+(1)}^{\mathbb R}\phi(x) = \int_{-\infty}^{\infty} dx\,
f(x)\left[\phi(x)-\phi(1)\right],
\label{eq:full-line-plus-app}
\end{equation}
where $\phi(x)$ is a smooth test function that decreases sufficiently fast at infinity. When there is no confusion, we write
\begin{equation}
\left[f(x)\right]_{+(1)} \equiv \left[f(x)\right]_{+(1)}^{\mathbb R}.
\end{equation}
With this convention,
\begin{equation}
\int_{-\infty}^{\infty} dx\, \left[f(x)\right]_{+(1)}^{\mathbb R}=0 ,
\label{eq:full-line-plus-zero}
\end{equation}
which is the basis of quark-number conservation for the corresponding one-loop quasi-PDF.

Equivalently, one may write
\begin{equation}
\left[f(x)\right]_{+(1)}^{\mathbb R} = f(x) - \delta(1-x) \int_{-\infty}^{\infty} dy\, f(y),
\label{eq:full-line-plus-delta}
\end{equation}
whenever the integral is defined with the same prescription.  This symbolic form is useful for keeping track of the 
$\delta(1-x)$ term, but Eq.~\eqref{eq:full-line-plus-app} is the defining equation.

\subsection{Regional plus prescriptions at $x=1$}

When the distribution on the full line is written separately in different regions, we use a superscript to indicate the domain 
of the plus prescription.  For the physical region $0<x<1$,
\begin{equation}
\int_0^1 dx\, \left[f(x)\right]_{+(1)}^{[0,1]}\phi(x) = \int_0^1 dx\, f(x)\left[\phi(x)-\phi(1)\right].
\label{eq:region-plus-01}
\end{equation}
Similarly, for the region $x>1$,
\begin{equation}
\int_1^\infty dx\, \left[f(x)\right]_{+(1)}^{[1,\infty)}\phi(x) = \int_1^\infty dx\, f(x)\left[\phi(x)-\phi(1)\right],
\label{eq:region-plus-1inf}
\end{equation}
provided the integral is convergent at infinity.  The notation $\left[\,\cdots\,\right]_{+(1)}^{D}$ always means that 
the subtraction point is the finite endpoint $x=1$, while the superscript $D$ specifies the domain on which the distribution 
is considered.

This notation should be distinguished from a plus prescription at infinity. For example,
\begin{equation}
\left[f(x)\right]_{+(1)}^{[1,\infty)}
\end{equation}
denotes a plus prescription at $x=1$ on the interval $[1,\infty)$, but it does not by itself prescribe the behavior at $x=\infty$.

\subsection{Boundary prescriptions at infinity}

If a function has a nonintegrable tail at infinity, a separate boundary prescription is required.  For the logarithmically 
divergent tail $1/x$ on $[a,\infty)$, we define
\begin{equation}
\int_a^\infty dx\, \left(\frac{1}{x}\right)_{+(\infty)}^{[a,\infty)} \phi(x) =
\int_a^\infty dx\,\frac{1}{x}\left[\phi(x)-\phi(\infty)\right],
\label{eq:plus-infinity}
\end{equation}
where
\begin{equation}
\phi(\infty)\equiv \lim_{x\to\infty}\phi(x)
\end{equation}
when this limit exists.  For ordinary test functions, $\phi(\infty)=0$, and the prescription reduces to the usual integral.  The
notation is nevertheless useful when boundary distributions at infinity are kept explicitly.

Similarly, on the negative half line we define
\begin{equation}
\int_{-\infty}^{a} dx\, \left(\frac{1}{1-x}\right)_{+(\infty)}^{(-\infty,a]} \phi(x) = \int_{-\infty}^{a} dx\,\frac{1}{1-x}
\left[\phi(x)-\phi(-\infty)\right],
\label{eq:plus-minus-infinity}
\end{equation}
with
\begin{equation}
\phi(-\infty)\equiv \lim_{x\to-\infty}\phi(x).
\end{equation}
The symbol $+(\infty)$ is used for prescriptions at infinity in the positive variables $x$ and $1-x$.  It should not be confused 
with the finite-endpoint prescription $+(1)$.

\subsection{Boundary delta functions}

It is sometimes convenient to represent boundary terms at infinity by delta
functions in the variable $1/x$.  We define
\begin{align}
\int_0^\infty dx\, \frac{1}{x^2}\delta^+\!\left(\frac{1}{x}\right)\phi(x) &= \phi(\infty), \nonumber \\
\int_{-\infty}^0 dx\, \frac{1}{(1-x)^2}\delta^+\!\left(\frac{1}{1-x}\right)\phi(x) &= \phi(-\infty).
\label{eq:delta-plus-negative-infinity}
\end{align}
Here $\delta^+(1/x)$ denotes the boundary delta function at $x=+\infty$, while $\delta^+(1/(1-x))$ denotes the 
boundary delta function at $x=-\infty$, written in terms of the positive variable $1-x$.  These distributions are useful for 
writing regulated large-$|x|$ limits in a form that makes the boundary contribution explicit.

For example, a regulated expression of the form $\theta(x-a)/x^{1+\eps}$ has the small-$\epsilon$ expansion
\begin{equation}
\frac{\theta(x-a)}{x^{1+\eps}} = -\frac{1}{\euv}\, \frac{1}{x^2}\delta^+\!\left(\frac{1}{x}\right) +
\left(\frac{1}{x}\right)_{+(\infty)}^{[a,\infty)} +\mathcal{O}(\eps),
\label{eq:large-x-expansion-plus}
\end{equation}
up to terms localized at the lower endpoint $x=a$ if that endpoint is also
singular.  Likewise,
\begin{equation}
\frac{\theta(a-x)}{(1-x)^{1+\eps}} = -\frac{1}{\euv}\, \frac{1}{(1-x)^2}\delta^+\!\left(\frac{1}{1-x}\right) +
\left(\frac{1}{1-x}\right)_{+(\infty)}^{(-\infty,a]} + \mathcal{O}(\epsilon).
\label{eq:large-x-expansion-minus}
\end{equation}
The sign and normalization follow from the definitions in Eq.~\eqref{eq:delta-plus-negative-infinity}.

\subsection{Application to the minimal quasi-PDF}

The function $h_{\mRI}(x,\rho)$ used in the main text is not obtained by deleting every term in $h_0(x,\rho)$ that is 
independent of the collinear logarithm. Rather, it is defined by keeping the minimal physical collinear structure while 
preserving the large-$|x|$ behavior required for a distribution on the full line.  Thus
\begin{equation}
h_{\mRI}(x,\rho)=
\begin{cases}
\displaystyle \frac{1+x^2}{1-x}\ln\frac{x}{x-1}+1, & x>1,\\[0.7em]
\displaystyle \frac{1+x^2}{1-x}\ln\frac{4}{\rho}, & 0<x<1,\\[0.7em]
\displaystyle \frac{1+x^2}{1-x}\ln\frac{x-1}{x}-1, & x<0 .
\end{cases}
\label{eq:hmin-app}
\end{equation}
The constants in the regions $x>1$ and $x<0$ are required because
\begin{align}
&\frac{1+x^2}{1-x}\ln\frac{x}{x-1}\longrightarrow -1, \qquad x\to+\infty, \nonumber \\
&\frac{1+x^2}{1-x}\ln\frac{x-1}{x}\longrightarrow +1, \qquad x\to-\infty.
\end{align}
Therefore the combinations appearing in Eq.~\eqref{eq:hmin-app} vanish at large $|x|$.
This large-$|x|$ completion is part of the definition of the minimal distribution on the full line.  The difference between the 
minimal scheme and the modified minimal scheme is then confined to the physical region:
\begin{equation}
h_0(x,\rho)-h_{\mRI}(x,\rho) = \begin{cases}
0, & x>1,\\[0.7em]
\displaystyle -\frac{2x}{1-x}, & 0<x<1,\\[0.7em]
0, & x<0 .
\end{cases}
\label{eq:h0-minus-hmin-app}
\end{equation}
Consequently,
\begin{equation}
H_0(x,\rho)-H_{\mRI}(x,\rho) = \left[-\frac{2x}{1-x}\right]_{+(1)}^{[0,1]} .
\label{eq:H0-minus-Hmin-app}
\end{equation}
This is the finite distribution that relates the matching coefficients in the minimal and modified minimal schemes.

\bibliographystyle{JHEP1}
\bibliography{scheme}

\end{document}